\documentclass[a4paper,fleqn,usenatbib]{mnras}

\usepackage[T1]{fontenc}
\usepackage{ae,aecompl}

\usepackage{graphicx}	
\usepackage{amsmath}	
\usepackage{amssymb}	
\usepackage{upgreek}

\title[Two-temperature magnetized accretion flow]{Properties of two-temperature magnetized advective accretion flow around rotating black hole}

\author[I. K. Dihingia et al.]{
Indu K. Dihingia,$^{1,2}$\thanks{E-mail: i.dihingia@iitg.ac.in}
Santabrata Das,$^{1}$\thanks{E-mail: sbdas@iitg.ac.in}
Geethu Prabhakar,$^{3}$\thanks{E-mail: geethuprabhakar.17@res.iist.ac.in}
and Samir Mandal$^3$\thanks{E-mail: samir@iist.ac.in}
\\
$^{1}$Indian Institute of Technology Guwahati, Guwahati 781039, Assam, India\\
$^{2}$Discipline of Astronomy, Astrophysics and Space Engineering, Indian Institute of Technology Indore, Indore 453552, India\\
$^3$Indian Institute of Space Science and Technology, Thiruvananthapuram, India\\
}

\date{Accepted XXX. Received YYY; in original form ZZZ}

\pubyear{2019}

\begin{document}
\label{firstpage}
\pagerange{\pageref{firstpage}--\pageref{lastpage}}
\maketitle

\begin{abstract}
    We study the two-temperature magnetized advective accretion flow around the Kerr black holes. During accretion, ions are heated up due to viscous dissipation, and when Coulomb coupling becomes effective, they transfer a part of their energy to the electrons.  On the contrary, electrons lose energy due to various radiative cooling processes, namely bremsstrahlung, synchrotron, and Comtonization processes, respectively. To account for the magnetic contribution inside the disc, we consider the toroidal magnetic fields which are assumed to be dominant over other components. Moreover, we adopt the relativistic equation of state to describe the thermal characteristics of the flow. With this, we calculate the global transonic accretion solutions around the rotating black holes. We find that accretion solution containing multiple critical points may harbor shock wave provided the standing shock conditions are satisfied. Further, we investigate the shock properties, such as shock location ($x_s$) and compression ratio ($R$) that delineate the post-shock corona (hereafter PSC) and find that the dynamics of PSC is controlled by the flow parameters, such as accretion rate (${\dot m}$) and magnetic fields ($\beta$, defined as the ratio of gas pressure to the magnetic pressure), etc. Finally, we calculate the emission spectra of the accretion flows containing PSC and indicate that both ${\dot m}$ and $\beta$ play the pivotal roles in explaining the spectral state transitions commonly observed for black hole X-ray binaries.
\end{abstract}


 \clearpage
\section{Introduction}

Black Hole X-ray binaries (BH-XRBs) are regarded as one of the fascinating objects in the sky which are in general seen to remain in the quiescent state for a long period followed by the occasional outbursts (e.g., GRO J1655-40, GS1354-64, GX339-4, SWIFTJ1745-26, V-404 Cyg, etc.). The spectral properties of the BH-XRBs are primarily classified into four major classes, namely, low-hard state (LHS), hard-intermediate state (HIMS), soft-intermediate state (SIMS) and high-soft state (HSS), respectively \citep{Belloni2010,Belloni-etal2011}. It is believed that the power-law component of the observed BH spectra is yielded because of the inverse Comptonization of the soft photons intercepted by the electrons of the hot corona at the inner part of the disc \citep{Chakrabarti-Titarchuk1995}. Hence, the temperature distribution of the hot corona, particularly the thermal properties of the coronal electrons, plays a pivotal role in deciding the spectral states of the accretion disc. Following this, several attempts were made in the theoretical front to investigate the spectral properties of the accretion disk using  \cite{Shakura-Sunyaev1973} standard disk model   \citep{Shields1978,Malkan-Sarget1982,Malkan1983,Madau1988,Laor-Netzer1989,Sun-Malkan1989}. Subsequently, \cite{Chakrabarti-Wiita1992} showed that when standing shocks are present at the inner part of the disc, the spectra from the active galactic nuclei (AGNs) modify significantly. Meanwhile,\cite{Chakrabarti-Titarchuk1995,Mandal-Chakrabarti2005} coarsely investigated the dependence of disk spectra on the properties of the post-shock corona (hereafter PSC) considering two-temperature advective accretion flow model. Further, \cite{Narayan-etal1996,Nakamura-etal1997};\cite{Esin-etal1997,Manmoto-etal1997,Quataert-Narayan1999,Yuan-etal2003,Oda-etal2012} also studied the spectral behavior of the black hole sources considering the two-temperature advective accretion flow. Note that these studies dealt with accretion solution that pass through a single critical point (more precisely inner critical point only), and completely ignored an important class of the solutions, such as multi-transonic advective accretion solutions \citep{Fukue1987,Chakrabarti1989,Das-etal2001a,Das-etal2003,Das2007}. What is more is that numerous general relativistic magnetohydrodynamics (GRMHD) simulation studies suggest that two-temperature accretion flow model is potentially viable to explain the radiative properties of active galactic nuclei (AGNs), such as Sagittarius A* and M87, etc. \citep[and references therein]{Ressler-etal2015,Sadowski-etal2017,Ryan-etal2018,Chael-etal2019}.

In an accretion process, subsonic flow from the outer edge of the disc starts accreting towards the central black hole with negligible velocity due to the influence of gravity. As the subsonic flow moves in, its velocity increases, and at some point, the flow becomes supersonic when its velocity exceeds the local sound speed. Such a point where flow smoothly changes its sonic state from subsonic to supersonic is commonly called a critical point. After crossing the critical point, supersonic flow continues to accrete because of the fatal attraction of gravity and finally enters into the black hole. However, depending on the input flow parameters, the supersonic flow may experience centrifugal repulsion against gravity, and when the repulsion is strong enough, it eventually triggers the discontinuous transition of the flow variables in the form of shock waves. At the shock, the flow jumps from supersonic to the subsonic branch, and thus all its kinetic energy is converted to the thermal energy there. In addition, due to the shock transition, post-shock flow feels density compression as well. Thus, the post-shock flow becomes hot and dense. After the shock, flow again moves inward and gradually picks up its velocity and ultimately enters into the black hole supersonically after crossing another critical point near the horizon. Accretion solutions of this kind contain more than one critical point, and hence, they are known as multi-transonic advective accretion solutions \citep{Fukue1987,Chakrabarti1989,Das-etal2001a,Das-etal2003, Das2007}. Needless to mention that numerous groups of researchers extensively studied the shock induced advective accretion solutions around black holes both theoretically as well as numerically \citep{Fukue1987,Chakrabarti1989,Yang-Kafatos1995,Molteni-etal96,
	Ryu-etal1997,Lu-etal1999,Becker-Kazanas01,Fukumura-Tsuruta2004,Nishikawa-2005,
	Das2007,Kumar-etal2013,Das-etal2014,Okuda-Das2015,Sukova-Janiuk2015,Fukumura-etal2016,Sarkar-Das16,Aktar-etal2017,Dihingia-etal2018,Kim-2018,Okuda-etal2019}.

Meanwhile, \cite{Dihingia-etal2015,Dihingia-etal2018a,Dihingia-etal2018b} calculated all possible solutions, including the multi-transonic one for two-temperature advective accretion flow (hereafter TAAF) and for the first time, they self-consistently obtained the shock induced TAAF solutions around Schwarzschild black holes. Further, they indicated that since the shocked accretion solutions are thermodynamically preferred, precise disc spectra would be obtained when TAAF solutions containing shock waves are used for spectrum calculation.  Accordingly, it appears that TAAF solutions are perhaps potentially viable to explain the spectral properties of the black hole sources. Recently, \cite{Sarkar-Chattopadhyay2019} also examined the two-temperature accretion flow around a non-rotating black hole.

Motivating with this, we extend our previous works \citep{Dihingia-etal2015} by relaxing several underlying assumptions. For example, in an accretion disc, the flow remains non-relativistic at a large distance, and it becomes relativistic as it approaches the horizon. Accordingly, for a two-temperature flow, the adiabatic indices for both electrons ($\gamma_e$) and ions ($\gamma_i$) are expected to vary as the flow accretes towards the black hole.  In fact, a thermally relativistic fluid will have the ratio of specific heats $\gamma=4/3$, and subsequently, a thermally non-relativistic fluid will have the ratio of specific heats $\gamma=5/3$. Hence, for TAAF, $\gamma_{e,i} \rightarrow 5/3$ at the outer edge of the disc whereas $\gamma_{e,i} \rightarrow 4/3$ at the vicinity of the black hole horizon \citep{Taub1948, Mignone-etal2005,Ryu-etal2006}. In order to incorporate the $\gamma_{e,i}$ variation, in the present work, we consider the relativistic equation of state (REoS) for ionized fluid \citep{Chandrasekhar1939,Synge1957}. This REoS evidently ensures that $\gamma_{e,i}$ continues to remain the explicit function of flow temperature.

Very recently, Dihingia et al. (2019) showed that the accretion flow structure changes significantly with the inclusion of physically motivated REoS instead of ideal equation of state (IEoS) as latter prescription assumes constant adiabatic index. In particular, they pointed out that there exists an upper bound of the location of the outer critical point when REoS is adopted and such bound does not exist when IEoS is used. This bound eventually enforces a limit on the maximum shock radius. Since shock properties seems to play decisive role in determining the disk spectra \cite[]{Chakrabarti-Titarchuk1995,Mandal-Chakrabarti2005}, accretion solutions obtained from REoS is expected to render physically consistent disk spectra.
	
It may be recall that most of the black hole sources, if not all, have non-zero spin value \citep{Shafee-etal06,Gou-etal2009,Aschenbach2010,Liu_etal2010,Gou-etal2011,Ludlam-etal15} and therefore, it is more appropriate to study the hydrodynamical properties of the accreting matter around rotating black holes. Keeping this in mind, in the present work, we use an effective potential for Kerr black hole which is recently developed by \cite{Dihingia-etal2018} (hereafter DDMC18 potential) and obtain the transonic accretion solutions. 
It is noteworthy that DDMC18 potential is derived from the first principle of the general relativistic hydrodynamics and therefore, this effective potential is free from any limitations due to the spin of the black hole unlike the other pseudo potentials available in the literature \citep{Chakrabarti-Khanna1992,Artemova-etal1996,Semerak-Karas1999,Mukhopadhyay2002,Ivanov-Prodanov2005,Chakrabarti-Mondal2006}. In obtaining the effective potential, \cite{Dihingia-etal2018} derives the relativistic Navier-Stokes equations in the co-rotating frame considering Kerr space-time and identifies the effective potential upon comparing it with the conventional Euler equation in the non-relativistic limit.

What is more, is that the magnetic field is ubiquitous in nature, and it must be present inside the viscous accretion disc also. Following the work of \cite{Oda-etal2007,Oda-etal2010}, in this work, we consider the azimuthally dominated structured magnetic fields and investigate the role of magnetic fields in determining the structure of the accretion disc around rotating black holes. In addition, we consider all the relevant radiative cooling mechanisms, namely bremsstrahlung, synchrotron, and Comtonization processes, which are active inside the disk. With this, we obtain the transonic as well as shocked accretion solutions around rotating black holes. Further, we study the shock properties, including shock radius and compression ratio, in terms of the flow parameters, namely accretion rate (${\dot m}$), viscosity ($\alpha_B$) and plasma $\beta$. Finally, we study the emission spectrum from an accretion disk and study the evolution of spectra with the flow parameters. Such spectral evolution leads to the variation of the photon index ($\Gamma$) which is directly linked with the spectral state transitions. With this, we infer the possible spectral state transition commonly seen in black hole sources using our model formalism.

The paper is organized in the following order. In Section 2, the model has been discussed. In Section 3, critical point analysis and solution methodology are described.  Subsequently, in Section 4, we present the results in detail. Finally, in Section 5, we summarize our findings.

\section{Model Equations and Assumptions}

\subsection{Basic Hydrodynamics}

We consider a steady, two-temperature, magnetized, axisymmetric advective accretion disc around a Kerr black hole. To express the flow variables, we use a unit system as $M_{\rm BH}=G=c=1$, where $M_{\rm BH}$ is the mass of the black hole, $G$ is the gravitational constant and $c$ is the speed of light, respectively. In this unit system, the radial coordinate, angular momentum, and flow velocity are measured in units of $GM_{\rm BH}/c^2$, $GM_{\rm BH}/c$, and $c$, respectively. Here, the mass of the black hole is chosen as $M_{\rm BH}=10M_{\odot}$ throughout the study, where $M_{\odot}$ represents the solar mass. Moreover, we assume that magnetic fields in the disc are turbulent in nature, and its azimuthal component dominates over other components  \citep{Machida-etal2006,Hirose-etal2006}. Hence, following \cite{Oda-etal2007}, we have the azimuthally averaged magnetic field as $\langle \vec{B}\rangle = \langle B_\phi \rangle \hat{\phi}$, where `$\langle ~ \rangle$' stands for the azimuthal average and $B_\phi$ denotes the azimuthal component of magnetic fields. In this work, we use a cylindrical coordinate system where the black hole resides at the origin of the coordinate system.

With the above considerations, we obtain the governing equations that describe the relativistic flow motion around the Kerr black hole \citep{Dihingia-etal2018} at the equatorial plane and are given by,

\noindent (a) Radial momentum equation:
$$
u\frac{du}{dx}+\frac{1}{h\rho}\frac{dP}{dx} 
+\frac{\langle B_{\phi}^2\rangle}{4\uppi\rho x} +
\frac{\partial \Psi_{\rm eff}}{\partial x}= 0,
\eqno(1)
$$
where $x$ is the radial distance, $u$ is the radial velocity, $\rho$ is the mass density and $h$ is the enthalpy. In equation (1), we assume Lorentz factor $\gamma_u \rightarrow 1$ as in general $u/c \lesssim 0.1$ for $r > 4GM_{\rm BH}/c^2$  \citep[references therein]{Dihingia-etal2018}. The isotropic total pressure is given by $P=P_{\rm gas}+P_{\rm mag}$, where $P_{\rm gas}$ is the gas pressure and $P_{\rm mag}$ is the magnetic pressure. The gas pressure ($P_{\rm gas}$) of the flow is essentially the sum of the partial pressures of the ions and electrons as
$P_{\rm gas}=\sum\limits_{j=i, e}\rho_j k_{\rm B} T_j/(\mu_j m_j),
$
where $k_{\rm B}$ is the Boltzmann constant, $T$ is the temperature in Kelvin and $i$ and $e$ refer the ions and electrons, respectively. With this, we have $h= (\epsilon + P_{\rm gas})/\rho$, where $\epsilon$ is the internal energy of the flow. In addition, $m_e$ and $m_i$ denote the mass of the electrons and ions, $k_{\rm B}$ is the Boltzmann constant, and $\mu_j$ are the mean molecular weights which are given by $\mu_{i}=1.23$ and $\mu_{e}=1.14$ for cosmic abundance of hydrogen mass fraction of $0.75$ \citep{Narayan-Yi1995}. The magnetic pressure inside the disc is obtained as $P_{\rm mag}= \langle B^2_\phi \rangle/8\pi$, where $\langle B^2_\phi \rangle$ represents the azimuthal average of the square of the toroidal component of the magnetic fields. Further, we define plasma $\beta=P_{\rm gas}/P_{\rm mag}$ that yields the total pressure as $P=P_{\rm gas}(1+1/\beta)$. The third and last terms in the left-hand side of equation (1) denote the magnetic tension force and effective force terms, respectively. Here, we adopt a pseudo-potential \citep{Dihingia-etal2018} that satisfactorily mimics the space-time geometry around a Kerr black hole on the equatorial plane and is given by,
$$
\Psi_{\rm eff}= \frac{1}{2} \ln\bigg[\frac{x\Delta}{a_{\rm k}^2 (x+2)-4 a_{\rm k} \lambda +x^3-\lambda ^2 (x-2)}\bigg],
$$
where $\lambda$ denotes the specific angular momentum of the flow and $\Delta=x^2 - 2x + a_{\rm k}^2$. We further define the rotational parameter of the black hole as  $a_{\rm k}=|J/M_{\rm BH}|$ with $J$ is the angular momentum of the black hole. In this work, for the purpose of representation, we consider that the black hole spin is aligned with the angular momentum of the accretion flow, however the alternate possibility exits when these vectors are misaligned.

\noindent (b) Mass conservation equation:
$$
\dot{M}=2\uppi u \Sigma \sqrt\Delta,
\eqno(2)
$$
where $\dot{M}$ represents the accretion rate that remains constant throughout the flow and $\Sigma$ refers the vertically integrated surface mass density of the accreting matter \citep{Oda-etal2010}. In the subsequent analysis, we express the accretion rate in terms of the Eddington accretion rate as ${\dot m}={\dot M}/{\dot M}_{\rm Edd}$, where $\dot{M}_{\rm Edd}=1.44\times10^{17}\left( \frac{M_{\rm BH}}{M_{\odot}}\right)$ g s$^{-1}$.

\noindent (c) Azimuthal momentum equation:
$$
u\frac{d\lambda(x)}{dx}+\frac{1}{\Sigma x}\frac{d}{dx}\left(x^2W_{x\phi}\right)=0,
\eqno(3)
$$
where we assume that the $x\phi$ component of the Maxwell stress, $W_{x\phi}$, is dominated over the other components of the vertically integrated total stress. Following \cite{Machida-etal2006}, we calculate $W_{x\phi}$ for flows having significant radial velocity \citep{Chakrabarti-Das2004} as
$$
W_{x\phi} = \frac{\langle B_x B_\phi\rangle}{4\pi}H(x)=-\alpha_B (W+\Sigma u^2),
\eqno(4)
$$
where $\alpha_B$ denotes the proportionality constant, and $W$ represents the vertically integrated pressure \citep{Oda-etal2010}. Following the work of \cite{Shakura-Sunyaev1973}, we treat $\alpha_B$ as a global parameter for the accretion flow, and when radial velocity becomes unimportant, equation (4) boils down to the seminal $\alpha$-viscosity prescription \citep{Shakura-Sunyaev1973}. In equation (4), $H(x)$ represent the half-thickness of the disc and following \citet{Riffert-Herold1995,Peitz_Appl1997}, we calculate $H(x)$ as 
$$
H=\sqrt{\left(P/\rho\right){\cal F}},
\eqno(5)
$$ 
where 
$$
\mathcal{F}=\frac{(x^2 + a_{\rm k}^2)^2 - 2\Delta a_{\rm k}^2}
{(x^2 + a_{\rm k}^2)^2 + 2\Delta a_{\rm k}^2}\left(1-\Omega\lambda\right)x^3,
$$
with $\Omega ~\left[=2 a_{\rm k}+\lambda  (x-2))/(a_{\rm k}^2 (x+2)-2 a_{\rm k} \lambda +x^3)\right]$ being the angular velocity of the flow.

\noindent (d) The advection equation of toroidal magnetic flux:

The advection rate of toroidal magnetic flux is described by the induction equation and its azimuthally averaged form is given by,

$$
\frac{\partial \langle B_\phi \rangle \hat{\phi}}{\partial t}=\nabla \times \left( \vec{u} \times \langle B_\phi \rangle \hat{\phi} -\frac{4\pi}{c}\eta \vec{j} \right),
$$
where $\vec{u}$, $\eta$, and $\vec{j}$ refer the velocity vector, resistivity, and current density of the flow, respectively. Because of the large length scale of the accretion disc, the Reynolds number is generally high, and hence, we neglect the magnetic-diffusion term. In addition, we ignore the dynamo term and further assume that the azimuthally averaged toroidal magnetic fields vanish at the disc surface. Considering all these, we obtain the advection rate of the toroidal flux in the steady-state as \citep{Oda-etal2007},
$$
{\dot \Phi}=-\sqrt{4\pi}uHB_0(x),
$$
where $B_0(x)$ is the azimuthally averaged toroidal magnetic field confined at the disc equatorial plane. Usually, in an accretion disc, ${\dot \Phi}$ is not conserved, and Global three-dimensional MHD simulation of \cite{Machida-etal2006} indicates that ${\dot \Phi}$ may inversely vary with radial coordinate. Following this, we set a relation ${\dot{\Phi}\propto x^{-\zeta}}$ \citep{Oda-etal2007}, where $\zeta$ is a parameter describing the magnetic flux advection rate. Finally, we obtain the parametric relation as
$$
\dot{\Phi}(x,\zeta,\dot{M})\equiv\dot{\Phi}_{\rm edge}(\dot{M})\left(\frac{x}{x_{\rm edge}}\right)^{-\zeta},
\eqno(6)
$$ 
where $\dot{\Phi}_{\rm edge}$ is the advection rate of toroidal magnetic flux calculated at the outer edge of the disc ($x_{\rm edge}$). In this work, for the purpose of representation, we choose $\zeta=1$ all throughout unless stated otherwise.

\noindent (e) The entropy equations for ions ($i$) and electrons ($e$):
$$
u\left(\frac{n_{i}}{\rho_{i}}\frac{dP_{i}}{dx}-\frac{(1 + n_i) P_i}
{\rho_i^2}\frac{d\rho_i}{dx}\right)=\Lambda_{i}-\Gamma_{i},
\eqno(7)
$$
and
$$
u\left(\frac{n_e}{\rho_e}\frac{dP_e}{dx}-\frac{(1 + n_e) P_e}
{\rho_e^2}\frac{d\rho_e}{dx}\right)=\Lambda_{e}-\Gamma_{e},
\eqno(8)
$$
where $n_{i,e}$ is the polytropic indices, $\Gamma_{i,e}$ is the dimensionless heating terms and $\Lambda_{i,e}$ is the dimensionless cooling terms, respectively. Note that, in order to obtain the heating and cooling terms in CGS unit, one requires to multiply a factor $c^5/(G M_{\rm BH})$ with them.

The ions are heated up due to viscous heating and its explicit expression is given by
\citep[and references therein]{Chakrabarti-Molteni1995,Dihingia-etal2018a}
$$
\Gamma_i=-\alpha_{\rm B} x\left(\frac{P}{\rho} + u^2\right)\frac{d\Omega}{dx},
\eqno(9)
$$
where $P~(=P_{\rm gas}+P_{\rm mag})$ is the total pressure of the flow. 

Ions generally cool down because of the transfer of energy from ions to electrons via  Coulomb coupling $(Q_{ei})$ and through the inverse bremsstrahlung
$(Q_{ib})$ process. Hence, the total cooling rate of ions is therefore given by,
$$
Q_i^- = Q_{ei} + Q_{ib},
\eqno(10)
$$
where the explicit expression of the Coulomb coupling in CGS unit ($i. e.$, in erg cm$^{-3}$ s$^{-1}$) is 
given by \citep{Stepney-Guilbert1983,Narayan-Yi1995}:
$$\begin{aligned}
&Q_{ei} = 5.61\times10^{-32} n^2 \frac{k_{\rm B}T_i-k_{\rm B}T_e}{K_2(1/\Theta_e)K_2(1/\Theta_i)}\times\\
&\left[ \frac{2(\Theta_e+\Theta_i)^2+1}{(\Theta_e+\Theta_i)}K_1\left(\frac{\Theta_e+\Theta_i}{\Theta_e\Theta_i}\right)+2K_0\left(\frac{\Theta_e+\Theta_i}{\Theta_e\Theta_i}\right)\right],
\end{aligned}\eqno(11)
$$
where $K_0$, $K_1$ and $K_2$ denote the modified Bessel functions, and $\Theta_i=k_{\rm B}T_i/m_{i}c^2$ and $\Theta_e = k_{\rm B}T_e/m_{e}c^2$. The inverse bremsstrahlung in CGS unit ($i. e.$, in erg cm$^{-3}$ s$^{-1}$) is given by 
\citep{Rybicki-Lightman1979,Colpi-etal1984,Mandal-Chakrabarti2005}: 
$$
Q_{ib} = 1.4\times10^{-27}n^2\left(\frac{m_e}{m_i}T_i\right)^{1/2},
\eqno(12)
$$
where $n$ denotes the number density distribution of the flow. Here, we assume that the ions are mostly protons and therefore, for simplicity,  the number density of ions and electrons are considered to be identical although their mean molecular weights are different. With this, we obtain the dimensionless total cooling
term for ions as
$\Lambda_i=(Q^{-}_i/\rho_i)\times (GM_{\rm BH}/c^5)$.

Interestingly, this Coulomb coupling acts as heating terms for electrons and hence, we get
$$
Q_e^+ = Q_{ei}.
\eqno(13)
$$

Since the mass of the electron is smaller than ion, electrons generally cool more efficiently by bremsstrahlung process $(Q_b)$, cyclo-synchrotron process $(Q_{cs})$, and Comptonization process $(Q_{mc})$, respectively. Thus, the overall electron cooling due to the above radiative processes is calculated as,
$$
Q_e^- = Q_b + Q_{cs} + Q_{mc}.
\eqno(14)
$$
The explicit expressions of these cooling terms in CGS unit ($i. e.$, in erg cm$^{-3}$ s$^{-1}$) are given below \citep{Rybicki-Lightman1979,Mandal-Chakrabarti2005}: 

\noindent (i) Bremsstrahlung process:
$$
Q_b = 1.4\times 10^{-27}n^2T_e^{1/2}\left(1 + 4.4\times10^{-10}T_e\right).
\eqno(15)
$$
\noindent (ii) Cyclo-synchrotron process:
$$
Q_{cs} = \frac{2\uppi}{3c^2}k_BT_e\frac{\nu_{a}^3}{x},
\eqno(16)
$$
where $\nu_{a}$ is the cut-off frequency of synchrotron self-absorption that depends on the electron temperature and the magnetic fields. Here, we consider only the thermal electrons to obtain $\nu_a$ by equating the surface and volume emission of synchrotron radiation \citep{Mandal-Chakrabarti2005}. The expression of $\nu_a$ is 
given by
$$
\nu_a = \frac{3}{2}\nu_0\Theta_e^2x_m,
\eqno(17)
$$
where 
$$
\nu_0 = \frac{eB}{2\uppi m_ec} \quad {\rm and} \quad x_m = \frac{2\nu}{3\nu_0\Theta_e^2}
\eqno(18)
$$
where $B$ denotes magnetic fields.

\noindent (iii) Comptonization process:
$$
Q_{mc} = Q_{cs}\mathbb{F},
\eqno(19)
$$
where $\mathbb{F}$ represents the enhancement factor for the Comptonization of synchrotron radiations. Following the prescription of \citet{Mandal-Chakrabarti2005}, in this work, we calculate $Q_{mc}$. At the end, we calculate the dimensionless total heating and cooling terms for electrons as:
$\Gamma_{e}=(Q_e^{+}/\rho_e)\times\left(GM_{\rm BH}/c^5\right)$ and 
$\Lambda_{e}=(Q_e^{-}/\rho_e)\times\left(GM_{\rm BH}/c^5\right)$, respectively.

The above set of governing equations is closed with an equation of state (EoS). Following \cite{Synge1957}, the relativistic EoS (REoS) of the flow containing electrons and ions can be written as,
$$
\epsilon=f\rho,
\eqno(20).
$$
where $\epsilon$ is the internal energy and
$$
f=\sum\limits_{j=i, e}\frac{k_j}{\mu_j}\left[\frac{\mu_j K_3\left(\mu_j/\Theta_j\right)}{K_2\left(\mu_j/\Theta_j\right)}-\Theta_j\right],
\eqno(21)
$$
with $K_3$ being the modified Bessel function. Moreover, here we define the sound speed as $a_{\rm s}=\sqrt{P_{\rm gas}/(\epsilon+P_{\rm gas})}$.

The polytropic index for ions and electrons are given by, 
$$
n_i = \frac{\mu_e}{k_e}\frac{\partial f}{\partial \Theta_e}, \qquad n_e = \frac{\mu_i}{k_i}\frac{\partial f}{\partial \Theta_i},
$$
where, $k_e=m_e/(m_e+m_i)$ and $k_i=m_i/(m_e+m_i)$, respectively. The explicit expressions of $n_e$ and $n_i$ are given in appendix-A.

The ratio of the specific heat corresponding to ions and electrons are calculated as,
$$
\gamma_i=1+\frac{1}{n_i}, \qquad \gamma_e=1+\frac{1}{n_e}.
$$

Using equations ($1-3$) and ($6-8$), we obtain the wind equation of the flow after some algebra (see appendix-B) which is given by,
$$
\frac{du}{dx}=\frac{\cal N}{\cal D},
\eqno(22)
$$
where $\cal N$ and $\cal D$ represent the numerator and denominator and their explicit expressions are given in appendix-B. Further, we express the derivatives of the other flow variables in terms of $(du/dx)$, which are given by,
$$
\frac{d\Theta_e}{dx}=\Theta_{e11}\frac{du}{dx}+\Theta_{e12},
\eqno(23)
$$
$$
\frac{d\Theta_i}{dx}=\Theta_{i11}\frac{du}{dx}+\Theta_{i12},
\eqno(24)
$$
$$
\frac{d\lambda}{dx}=\lambda_{11} \frac{du}{dx}+\lambda_{12},{\rm ~~and}
\eqno(25)
$$
$$
\frac{d\beta}{dx}=\beta_{11} \frac{du}{dx}+\beta_{12}.
\eqno(26)
$$
The explicit expression of the coefficients, namely $\Theta_{e11}$, $\Theta_{e12}$, $\Theta_{i11}$, $\Theta_{i12}$, $\lambda_{11}$, $\lambda_{12}$, $\beta_{11}$ and $\beta_{12}$ are given in appendix-B.

\section{Critical points and solution methodology}

Accretion flows around the black holes are necessarily transonic in nature and therefore, flow must pass through the critical point before falling into the black hole \citep{Chakrabarti1989,Lu-etal1999,Gu-Lu2002,Fukumura-Tsuruta2004,Le-Becker2005,Takahashi2007,Fukumura-Kazanas2007,Mukhopadhyay2008,Sinha-etal2009,Rajesh-Mukhopadhyay2010,Pu-etal2012,Das-Czerny2012,Agarwal-etal2012,Le-etal2016}. In order to obtain the accretion solution, it is a prevalent practice to start integration of the governing equations (22-26) from the critical point ($x_c$) \citep[and references therein]{Chakrabarti1989,Das-etal2001,Das2007,Dihingia-etal2019a,Dihingia-etal2019b}. At the critical point, both numerator ($\cal N$) and denominator ($\cal D$) of equation (22) vanishes simultaneously and thus the radial velocity gradient $(du/dx)$ at $x_c$ takes the $0/0$ form, $i.e.,$ $du/dx=0/0$. Since the velocity profile of the accretion flow remains smooth all throughout along the streamline (except at the shock transition which will be discussed in the subsequent section), we, therefore, apply the l{'}Hospital rule to calculate radial velocity gradient $(du/dx)_c$ at $x_c$. Usually, $(du/dx)_c$ assumes two values. When both values are real, and of opposite sign, it is called saddle type critical point \citep[and references therein]{Chakrabarti-Das2004}. For nodal type critical point, $(du/dx)_c$ are real and of same sign while for spiral type critical point, $(du/dx)_c$ are imaginary.
It may be noted that saddle type critical points are stable whereas spiral and nodal type critical points are unstable \citep{Kato-etal1993}, and therefore, saddle type critical points have special importance as the global transonic solutions only pass through them. Because of this, we, therefore, focus only on saddle type critical points that are hereafter referred to critical point. Depending on the flow parameters, accretion flow may possess more than one critical points. In such cases, when critical point forms close to the horizon, it is called as inner critical point ($x_{\rm in}$), and when it forms far away from the horizon, it is called as outer critical point ($x_{\rm out}$).  

\section{Results}

To calculate the accretion solution, one requires to solve the governing equations (22-26) by supplying a set of input parameters of the flow. Among these flow parameters, viscosity ($\alpha_B$) and accretion rate (${\dot m}$) are treated as the global parameters of the flow and the initial values of angular momentum ($\lambda_c$), plasma $\beta_c$ and electron temperature ($\Theta_{ec}$) at $x_c$ are used as the local flow parameters. Supplying all these flow parameters along with the black hole spin ($a_{\rm k}$), we solve equations, namely  ${\cal N}=0$ and ${\cal D}=0$ simultaneously to calculate radial velocity ($u_c$) and ion temperature ($\Theta_{ic}$) at $x_c$. Using the initial values of the flow variables at $x_c$, namely ($x_c, \lambda_c, \beta_c, \Theta_{ec}, \Theta_{ic}$) along with a given set of ($\alpha_B, {\dot m},a_{\rm k}$), we integrate equation (22): first, inwards up to the horizon and then, outwards up to the large distance, equivalently the outer edge of the disc ($x_{\rm edge}$). Finally, we join both parts of the solution to obtain the complete global transonic accretion solution around the rotating black holes.

\subsection{Global Accretion solution}

\begin{figure}
	\includegraphics[scale=0.4]{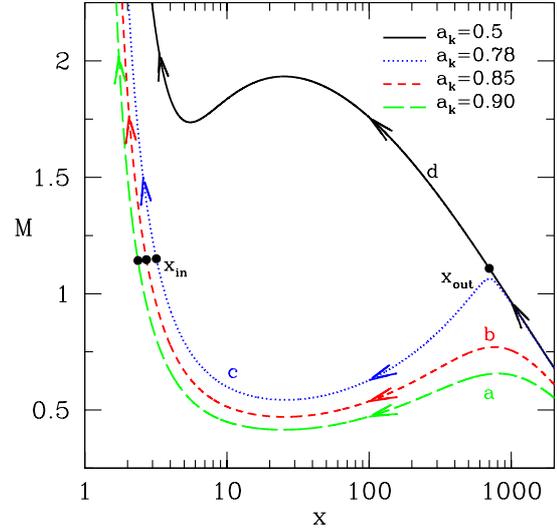}
	\caption{
		Typical accretion solutions where the variation of Mach number $(M=u/a_{\rm s})$ is plotted with radial distance $(x)$ for various Kerr parameter $(a_{\rm k})$. The letters a-d correspond to the different spin parameters marked in the figure. Filled circles (black) denote the location of the critical points which are marked. Arrows indicate the direction of flow the motion towards the black hole. See the text for details.
		}
\end{figure}

In Fig. 1, we show a typical example of accretion solution around a rotating black hole of spin $a_{\rm k}=0.90$, where Mach number ($M=u/a$) of the flow is plotted a function of radial distance. The solution depicted with big-dashed curve (in green) passes through the inner critical points $x_{\rm in}=2.3682$ with $\lambda_{\rm in}=2.0354$, $\beta_{\rm in}=13.6577$, $T_{e,~in}=8.1145\times10^{10}$K, $\alpha_B=0.01$ and ${\dot m}=0.001$ that connects the black hole horizon with the outer edge of the disk ($x_{\rm edge}=2000$). We mark this solution as `a'. For this solution, we note the values of the flow variables, namely $\alpha_B=0.01$, ${\dot m}=0.001$, $\lambda_{\rm edge}=2.6534$, $\beta_{\rm edge}=354.2302$, $u_{\rm edge}=0.008898$, $T_{i,~edge}=1.4790\times10^9{\rm K}$ and $T_{e,~edge}=1.2721\times10^9{\rm K}$ at $x_{\rm edge}=2000$. Indeed, one would get the identical accretion solution when the governing equations (22-26) are integrated from $x_{\rm edge}=2000$ to the horizon using the above outer edge flow variables. It may be noted that the above transonic accretion solution marked `a' is unique in nature which is obtained for a given set of input parameters. Next, we decrease the spin of the black hole and intend to obtain another unique transonic accretion solution keeping the flow parameters fixed at $x_{\rm edge}$. Now, since the spin value is changed, we need to change at least one flow variable at $x_{\rm edge}$ which we choose as $u_{\rm edge}$. This is advantageous over other flow variables in general as the spectral properties weakly depend on $u_{\rm edge}$. Following this, we decrease the black hole spin as $a_{\rm k}=0.85$ and inject the flow from $x_{\rm edge}=2000$ with the same flow variables as in case `a' except the radial velocity. We tune the radial velocity as $u_{\rm edge}=0.008898$ and obtain the global transonic accretion solution passes through the inner critical point $x_{\rm in}$. This solution is plotted using small dashed (in red) curve and marked as `b' in Fig. 1. Upon decreasing the black hole spin again as $a_{\rm k}=0.78$, we find the accretion solution `c' plotted with dotted curve (in blue) for $u_{\rm edge}=0.009910$, where the rest of the flow parameters at $x_{\rm edge}=2000$ remain same. Interestingly, as $a_{\rm k}$ is decreased further, accretion solution changes its character and passes through the outer critical point $x_{\rm out}$ instead of an inner critical point. Such a solution is depicted by the solid curve and marked as `d'. The solution of this kind has special importance as they may harbor shock wave (see \S 4.2) where infalling matter experiences discontinuous transitions in the flow variables. In the plot, inner and outer critical points are marked using the filled black circles, and the arrows indicate the direction of flow motion towards the black hole.

\subsection{Accretion solution with shock}

\begin{figure}
	\includegraphics[scale=0.4]{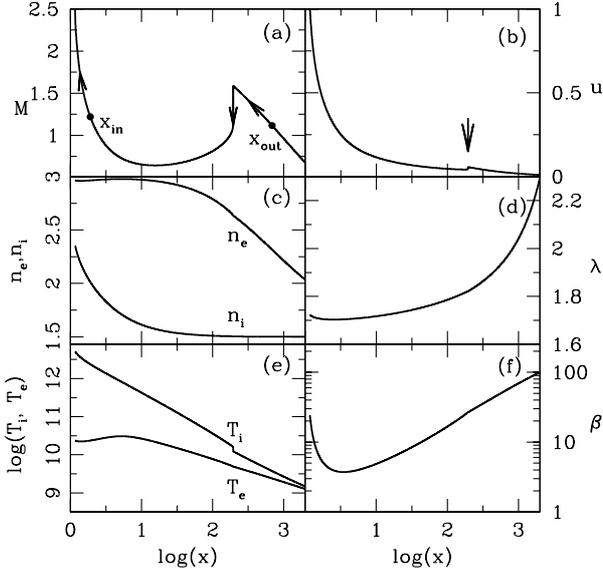}
	\caption{
		Variation of (a) Mach number $(M=u/a_{\rm s})$, (b) radial velocity $(u)$, (c) ion and electron polytropic indices $(n_i,n_e)$, (d) angular momentum $(\lambda)$, (e) ion and electron temperature $(T_{\rm i}, T_{\rm e})$, and (f) plasma $(\beta)$ as a function radial distance are depicted for accretion flow containing standing shock. The set of input parameters chosen here are identical to the result `d' of Fig. 1. See the text for details.
		}
\end{figure}

For accretion solution `d', subsonic flow first crosses the outer critical point at $x_{\rm out}=700$ to become supersonic and continues to proceed inward. Because of the centrifugal repulsion, inflowing matter slows down which eventually causes the accumulation of matter in the vicinity of the horizon. It results in a virtual barrier around the black hole where matter density is increased. Interestingly, the rise of density would not be continued indefinitely, and in some instances, the centrifugal barrier triggers the discontinuous transition of the flow variable in the form of shock wave provided the standing shock conditions are satisfied. In a vertically integrated flow, the conditions for shock transition are the conservation of (i) the mass flux: $\dot{m}_+=\dot{m}_-$ (ii) the energy flux: ${\cal E}_+={\cal E}_-$, (iii) the momentum flux: $W_{+} + \Sigma_{+} u_{+}^2=W_{-} + \Sigma_{-} u_{-}^2$, and (iv) the magnetic flux: ${\dot \Phi}_{+}={\dot \Phi}_{-}$ across the shock front, respectively where, the quantities with subscripts `$-$' and `$+$' refer their values immediately before and after the shock.  In condition (ii), ${\cal E}~[= u^2/2 + \log h + B_0^2/(4\pi \rho) +\Psi_{\rm eff}]$ represents the local Bernoulli parameter of the flow. In general, since the electron-electron collision time scale is much shorter than the ion-electron and ion-ion collision time scales \citep{Colpi-etal1984,Frank-etal2002}, we assume that the temperature profile of electrons remains unaffected across the shock transition and hence, we use $\Theta_{\rm e+}=\Theta_{\rm e-}$ across the shock front \citep{Dihingia-etal2018a}. Now, we describe the procedure to calculate the shock location employing the shock conditions stated above. At a virtual radius ($x^v$) of the accretion solution ($x^v < x_{\rm out}$) ($e. g.$, solution `d' in Fig. 1), we calculate the {\it shock conserved quantities} (i - iv) using the local supersonic flow variables, namely $u$, $\Theta_e$, $\Theta_i$, $\beta$, $\lambda$ and $\rho$. Following the method delineated by \citet{Chakrabarti-Das2004}, we utilize these {\it shock conserved quantities} to calculate the subsonic flow variables at $x^v$. Using these subsonic flow variables, we integrate equations (22-26) towards the event horizon to find the inner critical point ($x_{\rm in}$), where both ${\cal N}$ and ${\cal D}$ of equation (22) simultaneously tends to become zero. Once $x_{\rm in}$ is found, we further integrate equation (22-26) up to the event horizon in order to obtain the global shocked accretion solution where shock location is identified as $x_s = x^v$.	Upon failing to find $x_{\rm in}$, $x_v$ is updated by decreasing its value and the above procedure continues provided $dM/dx < 0 $ for the supersonic branch \citep[see][]{Das-etal2001a}. 

\begin{figure}
	\includegraphics[scale=0.4]{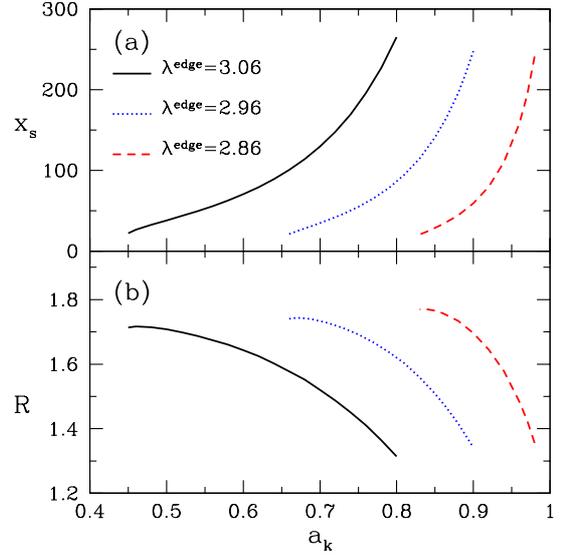}
	\caption{
		Variation of (a) shock location $(X_{\rm s})$ and (b) compression ratio $(R)$ as a function of Kerr parameter $(a_{\rm k})$ for accretion flows injected with three different angular momentum from $x_{\rm edge}=5000$. Dased (red), dotted (blue) and solid (black) curves denote the results for $\lambda_{\rm edge}=2.86$, $2.96$, and $3.06$, respectively. Here, we choose $\alpha_{\rm B}=0.01$, and $\dot{m}=0.01$. See the text for details.
		}
\end{figure}

To proceed further, we choose the solution `d' of Fig. 1 and find that shock conditions are satisfied at $x_s=122.4840$ where flow makes a discontinuous jump from the supersonic branch to the subsonic branch. Commonly, $x_s$ is called the shock location. After the shock, flow momentarily slows down and, thereafter, gradually picks up its radial velocity. Ultimately, the flow enters into the black hole supersonically after crossing the inner critical point at $x_{\rm in}=5.5618$. This shock-induced global accretion solution is shown in Fig. 2a, where critical points are marked, and the vertical arrow indicates the shock transition. In Fig. 2b, we show the radial velocity profile ($u$) of the accretion flow corresponding to the solution shown in Fig. 2a. Initially, $u$ increases monotonically up to the shock and drops off discontinuously at $x_s$. After the shock transition, $u$ again starts increasing and quickly accelerates in a way that flow reaches the event horizon with a velocity comparable to the speed of light. In Fig. 2c, we depict the polytropic indices of electrons ($n_e$) and ions ($n_i$) as a function of logarithmic radial distance. Here, we observe that electrons are always thermally relativistic all throughout, whereas ions mostly remain non-relativistic due to their high mass, except at the vicinity of the horizon. In Fig. 2d, the angular momentum profile of the flow is depicted. Due to viscosity, the angular momentum ($\lambda$) is transported outward, and we observe that flow approaches the event horizon with negligible $\lambda$. It is noteworthy that upon integrating equation (22-26) further out from $x_{\rm edge}$, $\lambda$ approaches the Keplerian angular momentum profile. Next, we present the temperature variation of electrons ($T_e$) and ions ($T_e$) in Kelvin in Fig. 2e. At $x_{\rm out}$, the electrons and ions are started with comparable temperatures. As the flow moves towards the horizon, radiative cooling processes become effective for electrons, while viscosity heats up the ions. Hence, the temperature profiles for ions and electrons start deviating from each other. At $x_s$, ions temperature shoots up discontinuously, although electron temperature remains smooth across the shock front as assumed. Due to shock compression, density, and temperature of the post-shock corona (PSC) are enhanced that results in the increase of the overall ion-electron temperature distributions at the PSC. Finally, we plot the variation of plasma-$\beta$ in Fig. 2f. We find that $\beta$ generally decreases with $x$. However, close to the horizon, gravitational compression increases due to the extreme curvature of the space-time and thus, the thermal pressure increases, which yield the increase of $\beta$ in the vicinity of the horizon.

Next, we examine the properties of shock waves. For that, we set $\alpha_B = 0.01$, ${\dot m}=0.01$ and inject accreting matter from $x_{\rm edge}=5000$ with $\beta_{\rm edge}=100$, $T_{e,~edge}=5.4987\times10^8{\rm K}$ and $T_{i,~edge}=6.6971\times10^8{\rm K}$, respectively. The obtained results are depicted in Fig. 3. In the upper panel (Fig. 3a), the variation of $x_s$ with $a_k$ is shown, where solid (black), dotted (blue) and dashed (red) curves are for $\lambda_{\rm edge}=3.06$, $2.96$ and $2.86$, respectively. For a given $\lambda_{\rm edge}$, $x_s$ recedes away from the horizon as $a_{\rm k}$ is increased. This gives a hint that the size of the post-shock corona (PSC) is enhanced with the increase of $a_{\rm k}$ for flows with an identical $\lambda_{\rm edge}$. Moreover, for a given $a_{\rm k}$, shock forms at larger radii when $\lambda_{\rm edge}$ is increased. This indicates that shocks are mainly centrifugally driven. Because of the shock transition, the post-shock flow is compressed, and the account of such compression is measured as the ratio of post-shock density to the pre-shock density, which is commonly called as the compression ratio ($R = \Sigma_{+}/\Sigma_{-}$). In the lower panel (Fig. 3b), we show the variation of $R$ with $a_{\rm k}$ for the same solutions presented in Fig. 3a. When the shock formation takes place at smaller radii, the gravitational potential energy released is higher that yields strong shock. Alternatively, as $a_{\rm k}$ is increased, shock moves away from the horizon, and compression ratio decreases that weakens the shock. It is to be noted that for a set of input parameters of the flow, there exists a range of $a_{\rm k}$, beyond which shock ceases to exist as shock conditions are not satisfied there.

\begin{figure}
	\includegraphics[scale=0.4]{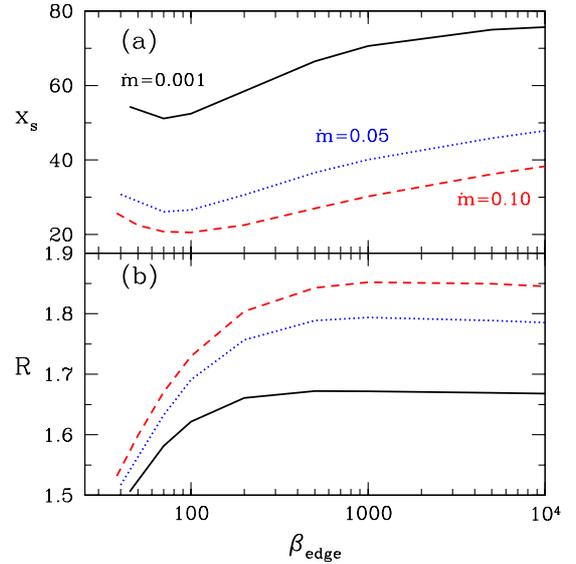}
	\caption{
		Variation of (a) shock location $(x_{s})$ and (b) compression ratio $(R)$ as a function of $\beta_{\rm edge}$ for flows injected with different accretion rate from $x_{\rm edge}=5000$. Solid (black), dotted (blue) and dashed (red) curves represent the results corresponding to $\dot{m}=0.001, 0.05$ and $0.1$, respectively. Here, we choose $a_{\rm k}=0.998$, and $\alpha_{\rm B}=0.01$. See the text for details.
		}
\end{figure}

In Fig. 4, we examine the impact of magnetic fields in the structure of accretion flow around a rotating black hole. As before, we choose
$\alpha_B = 0.01$, $\lambda_{\rm edge}=2.76$, $T_{e,~edge}=5.4987\times10^8{\rm K}$, $T_{i,~edge}=6.6971\times10^8{\rm K}$ and inject matter from $x_{\rm edge}=5000$ with different sets of (${\dot m},~\beta_{\rm edge}$) values. Here, we set $a_k=0.998$ and investigate the shock properties, which are depicted in Fig. 4. As before, in the upper panel, we present the variation of shock location $(x_s)$ as function of $\beta_{\rm edge}$. Solid (black), dotted (blue) and dashed (red) curves are for ${\dot m} = 0.001$, $0.05$ and $0.1$, respectively. As $\beta_{\rm edge}$ is decreased, the effect of magnetic activity inside the disc is increased that enhances the synchrotron emission, which in turn increases the amount of Comptonization as well. As a result, the post-shock flow cools down that reduces the post-shock pressure causing the shocks to settle down at lower radii. Interestingly, when $\beta_{\rm edge}$ is small ($\beta_{\rm edge}\lesssim 70$), the gas pressure and magnetic pressure at the inner part of the disk becomes comparable. In that scenarios, when $\beta_{\rm edge}$ is decreased, the magnetic pressure tends to dominate over the gas pressure, and because of this, post-shock pressure is enhanced that eventually pushed the shock front outwards as is being seen for $\beta_{\rm edge}\lesssim 70$ or so. If $\beta_{\rm edge}$ is decreased beyond its critical value, say $\beta^{\rm cri}_{\rm edge}$, the accretion solution passing through the outer critical point fails to reach the horizon and thus,  standing shock solutions cease to exist. It is noteworthy that the value of $\beta^{\rm cri}_{\rm edge}$ is not universal, as it depends on the other flow parameter as well. For instance, in figure (4), we observe that standing shock solutions sustain lower $\beta^{\rm cri}_{\rm edge}$ value when ${\dot m}$ is higher. In the lower panel (Fig. 4b), we plot the variation of the compression ratio ($R$) corresponding to the results shown in Fig. 4a. We find that $R$ generally remains insensitive when $\beta_{\rm edge}$ is large, however, $R$ decreases for flows with lower $\beta_{\rm edge}$ values.

\begin{figure}
	\includegraphics[scale=0.4]{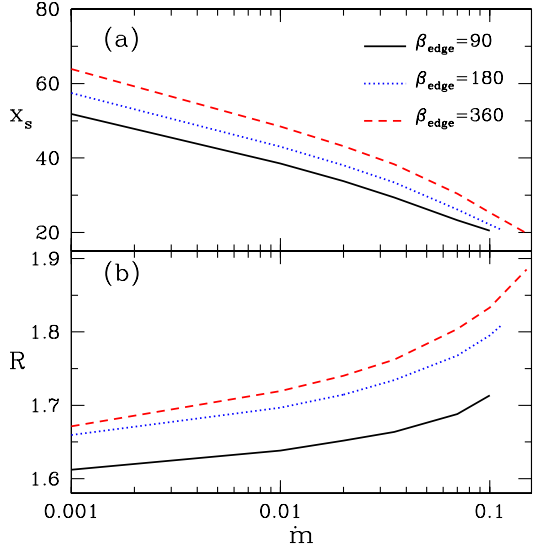}
	\caption{
		Variation of (a) shock location $(x_{s})$ and (b) compression ratio $(R)$ as a function of $\dot{m}$ for flows injected with different $\beta_{\rm edge}$ from $x_{\rm edge}=5000$. Solid (black), dotted (blue) and dashed (red) curves represent the results corresponding to $\beta_{\rm edge}=90, 180$ and $360$, respectively. Here, we choose $a_{\rm k}=0.998$, and $\alpha_{\rm B}=0.01$. See the text for details.
		}
\end{figure}

We continue the investigation of shock properties where the role of accretion rate $({\dot m})$ on the shock dynamics is studied. In Fig. 5, we show the variation of the shock location $(x_s)$ (in the upper panel) and compression ratio $(R)$ (in the lower panel) as a function of accretion rate (${\dot m}$). Here, we choose $a_{\rm k}=0.998$, $\alpha_B = 0.01$, $\lambda_{\rm edge}=2.76$, $T_{e,~edge}=5.4987\times10^8{\rm K}$, $T_{i,~edge}=6.6971\times10^8{\rm K}$ and inject matter from $x_{\rm edge}= 5000$ with three different $\beta_{\rm edge}$ values. The results depicted using solid (black), dotted (blue), and dashed (red) curves are obtained for $\beta_{\rm edge}=90$, $180$ and $360$, respectively. We find that $x_s$ is decreased with ${\dot m}$ for all the cases. In reality, as ${\dot m}$ is increased, the radiative processes become more effective, particularly at the PSC that reduces the thermal pressure there. Because of this, shock moves inwards and settles down at a smaller radius in order to maintain the pressure balance across it. On the other hand, for a given ${\dot m}$, as $\beta_{\rm edge}$ is decreased, the magnetic activity inside the disc is increased. This effectively enhances the synchrotron cooling, which in turn increases the Comptonization as well. As a consequence, the post-shock pressure is reduced due to the cooling down of the post-shock matter. This also compels the shock front to move inward to maintain the pressure balance across it. Overall, it is apparent that both $\beta_{\rm edge}$ and ${\dot m}$ concurrently play a decisive role in deciding the size of the PSC. What is more is that since the soft photons at PSC are further reprocessed by the hot electrons to produce hard radiation via the inverse-Comptonization process, it is perceived that the black hole spectral properties are expected to be regulated by ${\dot m}$ and $\beta_{\rm edge}$ as well.

\subsection{Accretion disk Spectrum}

\begin{figure}
	\includegraphics[scale=0.40]{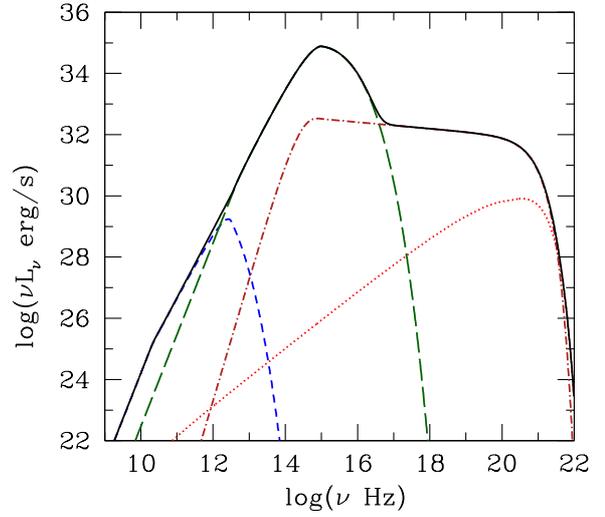}
	\caption{Plot of the typical emission spectrum for a shocked accretion solution depicted in Fig. 2. Dotted (red), dashed (blue), long-dashed (green), and the dot-dashed (magenta) curves represent the contribution of bremsstrahlung, pre-shock synchrotron, post-shock synchrotron, and Comptonization of the post-shock synchrotron photons, respectively and solid (black) curve denotes the total emission spectrum. Here, we consider $M_{\rm BH}=10~M_{\odot}$. See the text for details.
	}
\end{figure}

In this section, we calculate the typical spectrum of the shocked accretion flow around black holes. While doing this, we follow the works of \cite{Mandal-Chakrabarti2005} and \cite{Chakrabarti-Mandal2006}, where we take into the consideration of all the relevant contributions from the different radiative processes that are active inside the disc. For representation, we consider the shocked accretion solution presented in Fig. 2 and calculate the emission spectrum, which is depicted in Fig. 6. The different line styles denote the emission spectrum calculated for different radiative processes. For example, dotted (red), dashed (blue), long-dashed (green), and the dot-dashed (magenta) curves represent the contribution of bremsstrahlung, pre-shock synchrotron, post-shock synchrotron, and Comptonization of the post-shock synchrotron photons, respectively whereas solid (black) curve denotes the total emission spectrum. Here, the pre-shock synchrotron emission is calculated for all $x > x_s$ and post-shock synchrotron emission is calculated for all for $x < x_s$. Note that synchrotron photons constitute the lower energy part of the spectrum, and the Comptonization of the synchrotron photons constitutes the high energy tail. We calculate the photon index ($\Gamma$) in the frequency range $5\times 10^{18} - 5\times 10^{19}$ Hz that corresponds to energy range $20 - 200$ keV, and obtain as $\Gamma=2.11$.

\begin{figure}
	\includegraphics[scale=0.40]{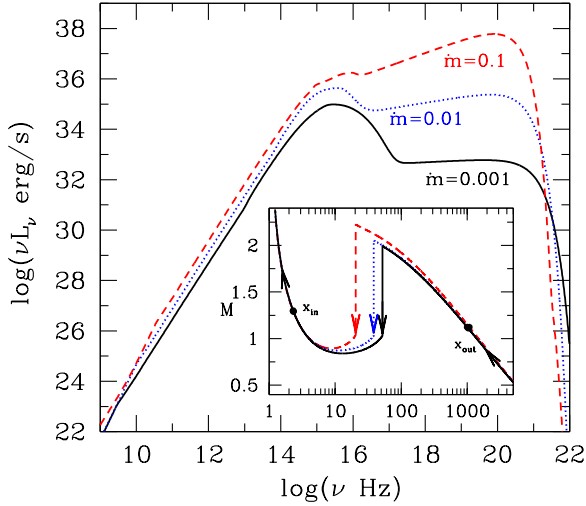}
	\caption{
		The modification of the emission spectrum with the accretion rate (${\dot m}$). Solid (black), dotted (blue) and dashed (red) spectra are for ${\dot m} = 0.001$, $0.01$ and $0.1$, respectively. Shocked accretion solutions that correspond to the spectra are shown at the inset using the identical line style.  Here, vertical arrows indicate the location of the shock transition and we consider $M_{\rm BH}=10~M_{\odot}$. See the text for details.
		}
\end{figure}

We further compare the disc emission spectra by considering three shocked accretion solutions, as shown in Fig. 7. Here, we choose $\alpha_B = 0.01$, $\lambda_{\rm edge}=2.76$, $\beta_{\rm edge}=90$, $T_{\rm e,edge}=5.4987\times10^8{\rm K}$, and ions temperature $T_{\rm i, edge}=6.6971\times10^8{\rm K}$ and inject matter from $x_{\rm edge}=5000$ with different ${\dot m}$. In the figure, spectra are drawn using solid (black), dotted (blue) and dashed (red) curves are for ${\dot m}=0.001$, $0.01$ and $0.1$, respectively. We calculate the photon index ($\Gamma$) for each spectrum and find that $\Gamma = 1.929$, $1.766$ and $1.282$ for ${\dot m}=0.001$, $0.01$ and $0.1$, respectively. At the inset of Fig. 7, the corresponding shocked accretion solutions are displayed for clarity. As ${\dot m}$ is increased, the availability of hot electrons at PSC is also increased, and hence, the radiative processes become efficient. This causes the cooling to be more intense, although PSC still remains hotter because of the supply of more hot electrons. With this, the overall luminosity increases. Due to all these, we observe that the calculated spectrum is appeared to be softer when ${\dot m}$ is small and gradually evolve to become hard for large ${\dot m}$.

\begin{figure}
	\includegraphics[scale=0.40]{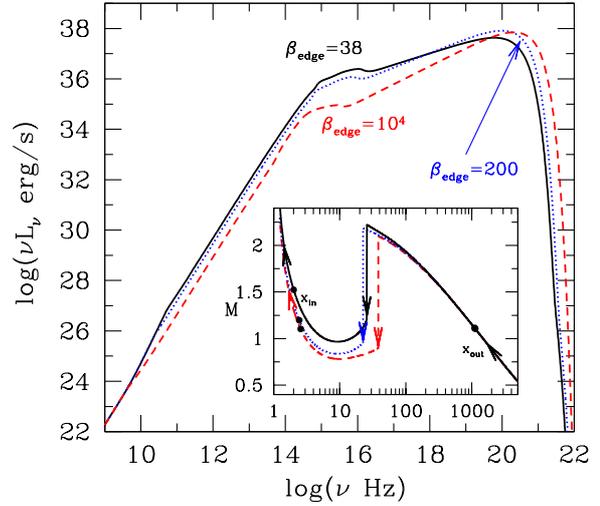}
	\caption{ 
		The modification of the emission spectrum with $\beta_{\rm edge}$. Solid (black), dotted (blue) and dashed (red) spectra are for $\beta_{\rm edge} = 38$, $200$ and $10000$, respectively. Shocked accretion solutions that correspond to the spectra are shown at the inset using the identical line style.  Here, vertical arrows indicate the location of the shock transition. We consider $M_{\rm BH}=10~M_{\odot}$. See the text for details.
	}
\end{figure}

In Fig. 8, we display the modification of the spectrum with the increase of the magnetic field strength. Here, we fix $\alpha_B = 0.01$, $\lambda_{\rm edge}=2.76$, ${\dot m}=0.1$, $T_{\rm e,edge}=5.4987\times10^8{\rm K}$ and $T_{\rm i, edge}=6.6971\times10^8{\rm K}$, and inject matter from $x_{\rm edge}=5000$ with different $\beta_{\rm edge}$.  Results plotted using solid (black), dotted (blue) and dashed (red) curves are for $\beta_{\rm edge}=38$, $200$ and $10^4$, respectively. The photon indices ($\Gamma$) calculated for the respective spectrum are $1.544$ (solid-black), $1.425$ (dotted-blue) and $1.282$ (dashed-red). For clarity, we show the corresponding shocked accretion solutions in the inset of Fig. 8. As $\beta_{\rm edge}$ increases, synchrotron cooling becomes inefficient which is indicated by the decrease of luminosity at $\nu \sim 10^{16} {\rm Hz}$. Hence, the spectrum becomes harder. We observe that the overall cooling increases for $\beta_{\rm edge}=200$, and the shock moves inward (blue-dotted vertical arrow) whereas for high $\beta_{\rm edge}$, shock moves outward (red-dashed vertical arrow) due to the significant reduction of the synchrotron cooling.

\begin{figure}
	\includegraphics[scale=0.40]{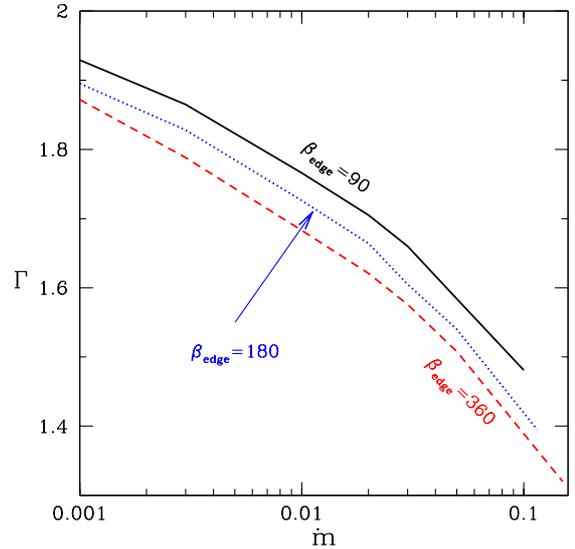}
	\caption{
		Variation of emission photon index $(\Gamma)$ with $\dot{m}$. Solid (black), dotted (blue) and dashed (red) curves are for $\beta_{\rm edge}=90$, $180$ and $360$, respectively. Usually, $1.4 \lesssim \Gamma \lesssim 1.8$ corresponds to hard state, $1.8 \lesssim \Gamma \lesssim 2.4$ for intermediate state and $2.4 \lesssim \Gamma \lesssim 3.5$ for soft state \citep[and references therein]{Nandi-etal2012}. See the text for details.
		}
\end{figure}

In general, the spectral states of BH-XRBs are identified with the photon index ($\Gamma$). Usually, the typical range of $\Gamma$ corresponding to the hard state is $1.4 \lesssim \Gamma \lesssim 1.8$, $1.8 \lesssim \Gamma \lesssim 2.4$ for intermediate state and $2.4 \lesssim \Gamma \lesssim 3.5$ for soft states, respectively \citep[and references therein]{Nandi-etal2012}. In order to facilitate that
in the next, we explore how the photon index ($\Gamma$) evolve with the input parameter of the accreting matter. While doing so, we present the variation of $\Gamma$ with ${\dot m}$ in Fig. 9. In this plot, the outer boundary parameters of the flow are chosen, as in Fig. 5. The results plotted using dashed (red), dotted (blue) and solid (black) curves are for $\beta_{\rm edge}=360$, $180$ and $90$, respectively. 
We observe that $\Gamma$ is decreased with ${\dot m}$ for all $\beta_{\rm edge}$. In general, for a given $\beta_{\rm edge}$, when ${\dot m}$ is increased, the overall density of the accretion flow is increased, and thus, the radiative cooling processes become efficient including the synchrotron emission process. Thus, plenty of synchrotron soft photons are available, which are being inverse Comptonized to produce hard photons. In a way, the energy of the hard photons is increased with an increase of ${\dot m}$, and hence, we find that the photon index ($\Gamma$) decreases with ${\dot m}$. On the other hand, for a given ${\dot m}$, $\Gamma$ increase with the decrease of $\beta_{\rm edge}$. This is expected as lower $\beta_{\rm edge}$ effectively weakens the efficiency of the Comptonization process resulting the increase of photon index ($\Gamma$). Further, it may be noted that, in this work, $\Gamma$ values remain restricted out side the range of the soft state. This is simply because, in our theoretical modeling, we neglect the contribution from the Keplerian disc at larger radii ($e.g.$, $x > x_{\rm edge}$) for simplicity and in order to obtain the soft state, an additional cooling of PSC due to soft X-ray thermal photons from Keplerian disc are required.

\section{Conclusion}

In this article, we investigate the properties of the two-temperature, magnetized, advective accretion flow around a rotating black hole. In order to take care of the general relativistic effect, we adopt DDMC18 potential \citep{Dihingia-etal2018} that satisfactorily describes the space-time geometry around the rotating black hole. In addition, we choose the relativistic equation of state (REoS) that delineates the thermodynamical characteristic of the accretion flow in a realistic manner. With this, we self-consistently solve the governing equations that describe the flow motion and obtain the transonic accretion solutions around a rotating black hole. We also find that depending on the input parameters, accretion flow may harbor standing shock wave provided the shock conditions are satisfied. Since the entropy content of the shocked accretion solution is higher than the shock free accretion solution \cite[]{Das-etal2001a}, according to the second law of thermodynamics, the shocked accretion solutions are preferred \cite[]{Becker-Kazanas01}, and such solutions are suitable to explain the spectral properties of the black hole sources \cite[]{Chakrabarti-Titarchuk1995,Mandal-Chakrabarti2005}. In the below, we summarize our findings based on the present work:

\begin{enumerate}
	    \item We find that depending on the input parameters, two-temperature magnetized advective accretion flow around rotating black hole possesses either single or multiple critical points (Fig. 1). 
	    
	    \item When multiple critical points exist, flow can pass through the shock wave provided the standing shock conditions are favorable. We obtain the shock-induced global accretion solutions and examine the accretion disc structure (Fig. 2).
	    
	    \item We investigate the shock properties, namely shock location ($x_s$) and compression ratio ($R$), by varying the input parameters and find that both $x_s$ and $R$ strongly depends on the input parameters (Fig. 3-5).
	    
	    \item We employ the two-temperature magnetized shocked accretion solution to calculate the emission spectrum (Fig. 6) and observe that accretion rate (${\dot m}$) (Fig. 7) and plasma $\beta$ (Fig. 8) play a crucial role in determining the emission spectra of the black hole sources.
	    
	    \item Finally, we study how ${\dot m}$ and plasma $\beta$ control the emission photon index ($\Gamma$). It turns out that emission spectra generally becomes harder ($i.e.$, $\Gamma$ decreases) when ${\dot m}$ and/or plasma $\beta$ are increased and {\it vice versa}. This evidently indicates that ${\dot m}$ and plasma $\beta$ seem to remain instrumental in explaining the black hole spectral state transitions. 
\end{enumerate}

	It is known that low mass X-ray binaries often display the spectral state transitions \citep{Homan-Belloni2005,Belloni-etal2006,Gierlinski-Newton2006,Miyakawa-etal2008}. 
	In particular, the {\it hysteresis} phenomenon is commonly seen in numerous astrophysical sources \citep{Klein-Wolt-etal2002,Kubota-Done2004,Rossi-etal2004,Zdziarski-etal2004,Meyer-Hofmeister2004,Dunn-etal2008,Obst-etal2013,Xiao-etal2018}. Here, we indicate that the present theoretical formalism seems to be viable in interpreting the observed spectral properties of the black hole sources by tuning the input flow parameters which we intend to take up as the future work and will be reported elsewhere.

\section*{Acknowledgments}

Authors thank the anonymous reviewer for constructive comments and suggestions that help to improve the quality of the manuscript. ID is indebted to Biplob Sarkar for discussion.



\appendix
\bsp	

\section{Equation of state}
The {\it equation of state} (EoS) given by \cite{Synge1957} for single component flow is written as,
$$
h=\frac{e+p}{\rho}=\frac{K_3(\rho/p)}{K_2(\rho/p)}~~~\implies e= \left[\frac{K_3(\rho/p)}{K_2(\rho/p)}-\frac{p}{\rho}\right]\rho,
$$
which can be rewritten for a flow with electrons and ions as,
$$
e=e_e+e_i=\rho f,
$$
where
$$
f=\frac{k_e}{\mu_e}\left[\frac{\mu_e K_3\left(\mu_e/\Theta_e\right)}{K_2\left(\mu_e/\Theta_e\right)}-\Theta_e\right]+ \frac{k_i}{\mu_i}\left[\frac{\mu_i K_3\left(\mu_i/\Theta_i\right)}{K_2\left(\mu_i/\Theta_i\right)}-\Theta_i\right].
$$
Here, the quantities have their usual meaning.
The polytropic indices are given as, $n_i = \frac{\mu_e}{k_e}\frac{\partial f}{\partial \Theta_e}$ and $n_e = \frac{\mu_i}{k_i}\frac{\partial f}{\partial \Theta_i}$ and their explicit expressions are given by,
$$
n_e=\frac{K_2\left(\frac{\mu_e}{\Theta_e}\right)^2+K_4\left(\frac{\mu_e}{\Theta_e}\right) K_2\left(\frac{\mu_e}{\Theta_e}\right)}{2 \left(\frac{\Theta_e}{\mu_e}\right)^2 K_2\left(\frac{\mu_e}{\Theta_e}\right)^2}
$$$$
-\frac{K_3\left(\frac{\mu_e}{\Theta_e}\right) \left[K_1\left(\frac{\mu_e}{\Theta_e}\right)+K_3\left(\frac{\mu_e}{\Theta_e}\right)\right]}{2 \left(\frac{\Theta_e}{\mu_e}\right)^2 K_2\left(\frac{\mu_e}{\Theta_e}\right)^2}-1
$$
$$
n_i=\frac{K_2\left(\frac{\mu_i}{\Theta_i}\right)^2+K_4\left(\frac{\mu_i}{\Theta_i}\right) K_2\left(\frac{\mu_i}{\Theta_i}\right)}{2 \left(\frac{\Theta_i}{\mu_i}\right)^2 K_2\left(\frac{\mu_i}{\Theta_i}\right)^2}
$$$$
-\frac{K_3\left(\frac{\mu_i}{\Theta_i}\right) \left[K_1\left(\frac{\mu_i}{\Theta_i}\right)+K_3\left(\frac{\mu_i}{\Theta_i}\right)\right]}{2 \left(\frac{\Theta_i}{\mu_i}\right)^2 K_2\left(\frac{\mu_i}{\Theta_i}\right)^2}-1.
$$
Here, $K_j$ ($j=1,2,3,4$) are the modified Bessel functions.

\onecolumn
\section{Calculation of Wind equation}
Using equation (2) in equations (1), (3), (6), (7), and (8) we obtain,
$$
R_0+ R_\beta \frac{d\beta}{dx}+ R_{\Theta_e} \frac{d\Theta_e}{dx}+R_{\Theta_i} \frac{d\Theta_i}{dx}+R_\lambda \frac{d\lambda}{dx}+R_u \frac{du}{dx}=0,
\eqno(A1)
$$
$$
L_0+ L_\beta \frac{d\beta}{dx}+ L_{\Theta_e} \frac{d\Theta_e}{dx}+L_{\Theta_i} \frac{d\Theta_i}{dx}+L_\lambda \frac{d\lambda}{dx}+L_u \frac{du}{dx}=0,
\eqno(A2)
$$
$$
B_0+B_{\beta}\frac{d\beta}{dx}+B_{\Theta_e} \frac{d\Theta_e}{dx} + B_{\Theta_i} \frac{d\Theta_i}{dx}+B_\lambda \frac{d\lambda}{dx}+ B_u \frac{du}{dx}=0
\eqno(A3)
$$
$$
E_0+ E_\beta \frac{d\beta}{dx}+ E_{\Theta_e} \frac{d\Theta_e}{dx}+E_{\Theta_i} \frac{d\Theta_i}{dx}+E_\lambda \frac{d\lambda}{dx}+E_u \frac{du}{dx}=0,
\eqno(A4)
$$
$$
P_0+ P_\beta \frac{d\beta}{dx}+ P_{\Theta_e} \frac{d\Theta_e}{dx}+P_{\Theta_i} \frac{d\Theta_i}{dx}+P_\lambda \frac{d\lambda}{dx}+P_u \frac{du}{dx}=0.
\eqno(A5)
$$
The coefficients of the equations (A1), (A2), (A3), (A4) and (A5) take the form

$$ 
\begin{aligned}
R_{0}=&\frac{\partial \Psi_{\rm eff}}{\partial x}+\frac{2 \left(\mu _e \Theta _i k_i+\Theta _e k_e \mu _i\right)}{\beta  x \mu _e \mu _i}+R_{01},~~ R_{01}=(\beta +1) \left(\mu _e \Theta _i k_i+\Theta _e k_e \mu _i\right)R_{02}/R_{03},\\
R_{02}=&-x \Delta  {\cal F}_0' \left(-4 \lambda  a_k+(x+2) a_k^2+x^3-\lambda ^2 (x-2)\right) \left(-2 \lambda  a_k+(x+2) a_k^2+x^3\right) + {\cal F}_0R_{04},\\
R_{03}=&2 \beta  x {\cal F}_0 \Delta  \left(-4 \lambda  a_k+(x+2) a_k^2+x^3-\lambda ^2 (x-2)\right) \left(-2 \lambda  a_k+(x+2) a_k^2+x^3\right) \left(\mu _i \left(\mu _e f+\Theta _e k_e\right)+\mu _e \Theta _i k_i\right),\\
R_{04}=&x \Delta ' \left(-2 \lambda  a_k+(x+2) a_k^2+x^3\right) \left(-4 \lambda  a_k+(x+2) a_k^2+x^3-\lambda ^2 (x-2)\right)\\
&+\Delta  \left(-\lambda ^2 \left((x (3 x+4)-36) a_k^2+x^4\right)-4 \lambda  \left(3 x^3 a_k+(4 x+9) a_k^3\right)+3 \left((x+2) a_k^2+x^3\right)^2+4 \lambda ^3 (2 x-3) a_k\right),\\
R_{u}=&u-\frac{(\beta +1) \left(\mu _e \Theta _i k_i+\Theta _e k_e \mu _i\right)}{\beta  u \left(\mu _i \left(\mu _e f+\Theta _e k_e\right)+\mu _e \Theta _i k_i\right)},~~
R_{\Theta_i}=\frac{(\beta +1) \mu _e k_i}{2 \beta  \left(\mu _i \left(\mu _e f+\Theta _e k_e\right)+\mu _e \Theta _i k_i\right)},~~ R_{\Theta_e}=\frac{(\beta +1) k_e \mu _i}{2 \beta  \left(\mu _i \left(\mu _e f+\Theta _e k_e\right)+\mu _e \Theta _i k_i\right)},\\
R_{\beta}=&-\frac{\mu _e \Theta _i k_i+\Theta _e k_e \mu _i}{2 \beta ^2 \left(\mu _i \left(\mu _e f+\Theta _e k_e\right)+\mu _e \Theta _i k_i\right)},~~ R_{\lambda}=\frac{(\beta +1) \left(\mu _e \Theta _i k_i+\Theta _e k_e \mu _i\right)}{\beta  \left(\mu _i \left(\mu _e f+\Theta _e k_e\right)+\mu _e \Theta _i k_i\right)}R_{\lambda 1},\\
R_{\lambda 1}=& \frac{a_k \left((x+2) a_k \left(a_k+\lambda  (x-2)\right)+x^3-\lambda ^2 (x-2)\right)+\lambda  (x-2) x^3}{\left(-4 \lambda  a_k+(x+2) a_k^2+x^3-\lambda ^2 (x-2)\right) \left(-2 \lambda  a_k+(x+2) a_k^2+x^3\right)},~~
{\cal F}_0=\frac{(x^2 + a_{\rm k}^2)^2 - 2\Delta a_{\rm k}^2}{(x^2 + a_{\rm k}^2)^2 + 2\Delta a_{\rm k}^2}, ~~{\cal F}_0'=\frac{\partial {\cal F}_0}{\partial x},~~\Delta'=\frac{\partial \Delta}{\partial x},\\
L_{0}=&-\frac{\alpha _B \left(4 \Delta -x \Delta '\right) \left((\beta +1) \mu _e \Theta _i k_i+\mu _i \left((\beta +1) \Theta _e k_e+\beta  u^2 \mu _e\right)\right)}{2 \beta  \Delta  \mu _e \mu _i},~~ L_{u}=\frac{x \alpha _B \left((\beta +1) \mu _e \Theta _i k_i+\mu _i \left((\beta +1) \Theta _e k_e-\beta  u^2 \mu _e\right)\right)}{\beta  u \mu _e \mu _i},\\
L_{\Theta_i}=&-\frac{(\beta +1) x \alpha _B k_i}{\beta  \mu _i},~~ L_{\Theta_e}=-\frac{(\beta +1) x \alpha _B k_e}{\beta  \mu _e},~~ L_{\beta}=\frac{x \alpha _B \left(\mu _e \Theta _i k_i+\Theta _e k_e \mu _i\right)}{\beta ^2 \mu _e \mu _i},\\
B_0=&\frac{x \Delta B_{01}{\cal F}_0'+ {\cal F}_0 B_{02}}{x {\cal F}_0 B_{04}}, ~~
B_{01}=\left(a^2 (x+2)-2 a \lambda +x^3\right) \left(a^2 (x+2)-\lambda  (4 a+(x-2) \lambda )+x^3\right),\\
B_{02}=&x \left(a^2 (x+2)-2 a_k \lambda +x^3\right)\left(a_k^2 (x+2)-\lambda  (4 a_k+(x-2) \lambda )+x^3\right) \Delta '+\Delta  B_{03},\\
B_{03}=&-4 \lambda  a_k \left(a_k^2 (12 \zeta +(6 \zeta +4) x+9)+3 (2 \zeta +1) x^3\right)\\&+(4 \zeta +3) \left((x+2) a_k^2+x^3\right)^2-\lambda ^2 \left(a_k^2 \left(-12 (4 \zeta +3)+(4 \zeta +3) x^2+4 x\right)+x^3 (-8 \zeta +4 \zeta  x+x)\right)+4 \lambda ^3 a_k (-4 \zeta +2 (\zeta +1) x-3),\\
B_{04}=&\Delta  \left(-4 \lambda  a_k+(x+2) a_k^2+x^3-\lambda ^2 (x-2)\right) \left(-2 \lambda  a_k+(x+2) a_k^2+x^3\right),\\
B_{\beta}=&\frac{2 \beta +3}{(1+\beta )\beta },~~
B_{\Theta_i}=-\frac{3 \mu _e k_i}{\mu _e \Theta _i k_i+\Theta _e k_e \mu _i},~~ B_{\Theta_e}=-\frac{3 k_e \mu _i}{\mu _e \Theta _i k_i+\Theta _e k_e \mu _i},\\
B_{\lambda}=&\frac{4 a_k+2 \lambda  (x-2)}{-4 \lambda  a_k+(x+2) a_k^2+x^3-\lambda ^2 (x-2)}
-\frac{2 a_k}{-2 \lambda  a_k+(x+2) a_k^2+x^3},~~ B_{u}=-\frac{2}{u},\\
E_{0}=&\frac{u \Theta _e}{2} \left(\frac{2 \left(2 a_k \left(\lambda -a_k\right)+(x-3) x^2\right)}{(x-2) \left(-\lambda  \left(4 a_k+\lambda  (x-2)\right)+(x+2) a_k^2+x^3\right)}-\frac{a_k^2+3 x^2}{-2 \lambda  a_k+(x+2) a_k^2+x^3}+\frac{\Delta '}{\Delta }-\frac{{\cal F}_0'}{{\cal F}_0}+\frac{1}{x-2}+\frac{3}{x}\right)-\frac{\mu _e}{k_e}\left(\Lambda_e-\Gamma_e\right),\\
E_{u}=&\Theta_e,~~ E_{\Theta_i}=\frac{u \Theta _e \mu _e k_i}{2 \mu _e \Theta _i k_i+2 \Theta _e k_e \mu _i},~~ E_{\Theta_e}=u \left(\frac{\Theta _e k_e \mu _i}{2 \mu _e \Theta _i k_i+2 \Theta _e k_e \mu _i}+n_e\right),~~
E_{\beta}=-\frac{u \Theta _e}{2 \beta  (\beta +1)},\\
E_{\lambda}=&\frac{u \Theta _e \left(-\lambda  (x-2) \left(-\lambda  a_k+(x+2) a_k^2+x^3\right)-a_k \left((x+2) a_k^2+x^3\right)\right)}{\left(-2 \lambda  a_k+(x+2) a_k^2+x^3\right) \left(-\lambda  \left(4 a_k+\lambda  (x-2)\right)+(x+2) a_k^2+x^3\right)}
-\frac{x^3 \left(a_k^2+(x-2) x\right) \alpha _B \left((\beta +1) (k_e \mu_i \Theta_e+ k_i \mu_e \Theta_i)+\mu_e \mu_i \beta  u^2\right)}{k_i \mu_e \beta\left(a_k^2 (x+2)-2 a_k \lambda +rx^3\right)^2},\\
P_{0}=&\frac{2 \alpha _B x \left(\lambda  \left(\lambda  a_k-2 a_k^2+(x-3) x^2\right)+3 x^2 a_k+a_k^3\right) \left((\beta +1) \left(\mu _e \Theta _i k_i+\Theta _e k_e \mu _i\right)+\beta  u^2 \mu _e \mu _i\right)}{\beta  \mu _e k_i \left(-2 \lambda  a_k+(x+2) a_k^2+x^3\right)^2},\\
&+\frac{u \Theta _i}{2} \left(-\frac{a_k^2+3 x^2}{-2 \lambda  a_k+(x+2) a_k^2+x^3}+\frac{2 \left(2 \lambda  a_k-2 a_k^2+(x-3) x^2\right)}{(x-2) \left(-\lambda  \left(4 a_k+\lambda  (x-2)\right)+(x+2) a_k^2+x^3\right)}+\frac{\Delta '}{\Delta }-\frac{{\cal F}_0'}{{\cal F}_0}+\frac{1}{x-2}+\frac{3}{x}\right)-\frac{\mu_i}{k_i}\Lambda_i,\\
\end{aligned}
$$

\clearpage

$$\begin{aligned}
P_{\Theta_i}=&u \left(\frac{\mu _e \Theta _i k_i}{2 \mu _e \Theta _i k_i+2 \Theta _e k_e \mu _i}+n_i\right),~~ P_{\Theta_e}=\frac{u k_e \Theta _i \mu _i}{2 \mu _e \Theta _i k_i+2 \Theta _e k_e \mu _i},~~
P_{\beta}=-\frac{u \Theta _i}{2 \beta  (\beta +1)},~~ P_{\lambda}=-\frac{P_{\lambda 0}}{\left(-2 \lambda  a_k+(x+2) a_k^2+x^3\right)^2}.\\
P_{\lambda 0}=&\frac{x^2 \alpha _B \left(a_k^2+(x-2) x\right) \left((\beta +1) \mu _e \Theta _i k_i+\mu _i \left((\beta +1) \Theta _e k_e+\beta  u^2 \mu _e\right)\right)}{\beta  \mu _e k_i},\\
&+\frac{u \Theta _i \left(-2 \lambda  a_k+(x+2) a_k^2+x^3\right) \left(\lambda  (x-2) \left(-\lambda  a_k+(x+2) a_k^2+x^3\right)+a_k \left((x+2) a_k^2+x^3\right)\right)}{-\lambda  \left(4 a_k+\lambda  (x-2)\right)+(x+2) a_k^2+x^3}.
\end{aligned}$$

With the help of equations (A1), (A2), (A3), (A4) and (A5), the wind equation is obtained as
$$
\frac{du}{dx}=\frac{\cal N}{\cal D},
\eqno(A6)
$$
where
$$
{\cal N}=-({\cal D}_1 R_0+{\cal N}_{10} R_{\Theta_e}+{\cal  N}_{20} R_{\Theta_i}+{\cal  N}_{30} R_\lambda +{\cal N}_{40} R_\beta),~~
{\rm and}~~
{\cal D}={\cal D}_1 R_u+{\cal N}_{1u} R_{\Theta_e}+{\cal  N}_{2u} R_{\Theta_i}+{\cal  N}_{3u} R_\lambda+{\cal N}_{4u} R_\beta.
$$
Here, 
$$\begin{aligned}
{\cal D}_1=&B_{\lambda } P_{\beta } E_{\Theta _i} L_{\Theta _e}-B_{\lambda } E_{\Theta _e} P_{\beta } L_{\Theta _i}-B_{\beta } P_{\lambda } E_{\Theta _i} L_{\Theta _e}+B_{\beta } E_{\Theta _e} P_{\lambda } L_{\Theta _i}-B_{\lambda } L_{\beta } E_{\Theta _i} P_{\Theta _e}+B_{\beta } L_{\lambda } E_{\Theta _i} P_{\Theta _e}+B_{\lambda } E_{\beta } P_{\Theta _e} L_{\Theta _i}-B_{\beta } E_{\lambda } P_{\Theta _e} L_{\Theta _i}\\
&+B_{\Theta _i} \left(L_{\Theta _e} \left(E_{\beta } P_{\lambda }-E_{\lambda } P_{\beta }\right)+E_{\Theta _e} \left(L_{\lambda } P_{\beta }-L_{\beta } P_{\lambda }\right)+P_{\Theta _e} \left(E_{\lambda } L_{\beta }-E_{\beta } L_{\lambda }\right)\right)+P_{\Theta _i} \left(E_{\Theta _e} \left(B_{\lambda } L_{\beta }-B_{\beta } L_{\lambda }\right)+L_{\Theta _e} \left(B_{\beta } E_{\lambda }-B_{\lambda } E_{\beta }\right)\right)\\
&+B_{\Theta _e} \left(E_{\Theta _i} \left(L_{\beta } P_{\lambda }-L_{\lambda } P_{\beta }\right)+L_{\Theta _i} \left(E_{\lambda } P_{\beta }-E_{\beta } P_{\lambda }\right)+P_{\Theta _i} \left(E_{\beta } L_{\lambda }-E_{\lambda } L_{\beta }\right)\right),\\
{\cal N}_{1u}=&\frac{1}{{\cal D}_2}\bigg[\left(B_{\lambda } \left(L_{\beta } P_u-L_u P_{\beta }\right)+B_{\beta } \left(L_u P_{\lambda }-L_{\lambda } P_u\right)+B_u \left(L_{\lambda } P_{\beta }-L_{\beta } P_{\lambda }\right)\right) \\
&\times\left(\left(B_{\beta } L_{\lambda }-B_{\lambda } L_{\beta }\right) \left(B_{\beta } E_{\Theta _i}-E_{\beta } B_{\Theta _i}\right)+\left(B_{\lambda } E_{\beta }-B_{\beta } E_{\lambda }\right) \left(B_{\beta } L_{\Theta _i}-L_{\beta } B_{\Theta _i}\right)\right)\bigg]\\
&+\frac{1}{{\cal D}_2}\bigg[B_{\beta } \left(B_{\lambda } \left(E_u L_{\beta }-E_{\beta } L_u\right)+B_u \left(E_{\beta } L_{\lambda }-E_{\lambda } L_{\beta }\right)+B_{\beta } \left(E_{\lambda } L_u-E_u L_{\lambda }\right)\right) \\
&\times\left(L_{\Theta _i} \left(B_{\beta } P_{\lambda }-B_{\lambda } P_{\beta }\right)+B_{\Theta _i} \left(L_{\lambda } P_{\beta }-L_{\beta } P_{\lambda }\right)+P_{\Theta _i} \left(B_{\lambda } L_{\beta }-B_{\beta } L_{\lambda }\right)\right)\bigg],\\
{\cal N}_{10}=&\frac{1}{{\cal D}_2}\bigg[\left(B_{\lambda } \left(P_0 L_{\beta }-L_0 P_{\beta }\right)+B_{\beta } \left(L_0 P_{\lambda }-P_0 L_{\lambda }\right)+B_0 \left(L_{\lambda } P_{\beta }-L_{\beta } P_{\lambda }\right)\right)\\
&\times \left(\left(B_{\beta } L_{\lambda }-B_{\lambda } L_{\beta }\right) \left(B_{\beta } E_{\Theta _i}-E_{\beta } B_{\Theta _i}\right)+\left(B_{\lambda } E_{\beta }-B_{\beta } E_{\lambda }\right) \left(B_{\beta } L_{\Theta _i}-L_{\beta } B_{\Theta _i}\right)\right)\bigg]\\
&+\frac{1}{{\cal D}_2}\bigg[B_{\beta } \left(B_{\lambda } \left(E_0 L_{\beta }-L_0 E_{\beta }\right)+B_0 \left(E_{\beta } L_{\lambda }-E_{\lambda } L_{\beta }\right)+B_{\beta } \left(L_0 E_{\lambda }-E_0 L_{\lambda }\right)\right)\\
&\times \left(L_{\Theta _i} \left(B_{\beta } P_{\lambda }-B_{\lambda } P_{\beta }\right)+B_{\Theta _i} \left(L_{\lambda } P_{\beta }-L_{\beta } P_{\lambda }\right)+P_{\Theta _i} \left(B_{\lambda } L_{\beta }-B_{\beta } L_{\lambda }\right)\right)\bigg],\\
{\cal N}_{2u}=&B_{\beta } E_{\Theta _e} L_{\lambda } P_u-B_{\beta } E_{\lambda } P_u L_{\Theta _e}-B_u E_{\Theta _e} L_{\lambda } P_{\beta }+B_u E_{\lambda } P_{\beta } L_{\Theta _e}-B_{\beta } E_{\Theta _e} L_u P_{\lambda }+B_u E_{\Theta _e} L_{\beta } P_{\lambda }+B_{\beta } E_u P_{\lambda } L_{\Theta _e}-B_u E_{\beta } P_{\lambda } L_{\Theta _e}\\
&+B_{\Theta _e} \left(E_{\lambda } \left(L_{\beta } P_u-L_u P_{\beta }\right)+E_{\beta } \left(L_u P_{\lambda }-L_{\lambda } P_u\right)+E_u \left(L_{\lambda } P_{\beta }-L_{\beta } P_{\lambda }\right)\right)+P_{\Theta _e} \left(B_u \left(E_{\beta } L_{\lambda }-E_{\lambda } L_{\beta }\right)+B_{\beta } \left(E_{\lambda } L_u-E_u L_{\lambda }\right)\right)\\
&+B_{\lambda } \left(E_{\Theta _e} \left(L_u P_{\beta }-L_{\beta } P_u\right)+L_{\Theta _e} \left(E_{\beta } P_u-E_u P_{\beta }\right)+P_{\Theta _e} \left(E_u L_{\beta }-E_{\beta } L_u\right)\right),\\
{\cal N}_{20}=&P_0 B_{\beta } E_{\Theta _e} L_{\lambda }-P_0 B_{\beta } E_{\lambda } L_{\Theta _e}-B_0 E_{\Theta _e} L_{\lambda } P_{\beta }+B_0 E_{\lambda } P_{\beta } L_{\Theta _e}-L_0 B_{\beta } E_{\Theta _e} P_{\lambda }+B_0 E_{\Theta _e} L_{\beta } P_{\lambda }+E_0 B_{\beta } P_{\lambda } L_{\Theta _e}-B_0 E_{\beta } P_{\lambda } L_{\Theta _e}\\
&+B_{\Theta _e} \left(E_{\lambda } \left(P_0 L_{\beta }-L_0 P_{\beta }\right)+E_{\beta } \left(L_0 P_{\lambda }-P_0 L_{\lambda }\right)+E_0 \left(L_{\lambda } P_{\beta }-L_{\beta } P_{\lambda }\right)\right)+P_{\Theta _e} \left(B_0 \left(E_{\beta } L_{\lambda }-E_{\lambda } L_{\beta }\right)+B_{\beta } \left(L_0 E_{\lambda }-E_0 L_{\lambda }\right)\right)\\
&+B_{\lambda } \left(E_{\Theta _e} \left(L_0 P_{\beta }-P_0 L_{\beta }\right)+L_{\Theta _e} \left(P_0 E_{\beta }-E_0 P_{\beta }\right)+\left(E_0 L_{\beta }-L_0 E_{\beta }\right) P_{\Theta _e}\right),\\
{\cal N}_{3u}=&B_{\beta } P_u E_{\Theta _i} L_{\Theta _e}-B_{\beta } E_{\Theta _e} P_u L_{\Theta _i}-B_u P_{\beta } E_{\Theta _i} L_{\Theta _e}+B_u E_{\Theta _e} P_{\beta } L_{\Theta _i}-B_{\beta } L_u E_{\Theta _i} P_{\Theta _e}+B_u L_{\beta } E_{\Theta _i} P_{\Theta _e}+B_{\beta } E_u P_{\Theta _e} L_{\Theta _i}-B_u E_{\beta } P_{\Theta _e} L_{\Theta _i}\\
&+B_{\Theta _i} \left(L_{\Theta _e} \left(E_u P_{\beta }-E_{\beta } P_u\right)+E_{\Theta _e} \left(L_{\beta } P_u-L_u P_{\beta }\right)+P_{\Theta _e} \left(E_{\beta } L_u-E_u L_{\beta }\right)\right)+P_{\Theta _i} \left(E_{\Theta _e} \left(B_{\beta } L_u-B_u L_{\beta }\right)+L_{\Theta _e} \left(B_u E_{\beta }-B_{\beta } E_u\right)\right)\\
&+B_{\Theta _e} \left(E_{\Theta _i} \left(L_u P_{\beta }-L_{\beta } P_u\right)+L_{\Theta _i} \left(E_{\beta } P_u-E_u P_{\beta }\right)+P_{\Theta _i} \left(E_u L_{\beta }-E_{\beta } L_u\right)\right),\\
{\cal N}_{30}=&P_0 B_{\beta } E_{\Theta _i} L_{\Theta _e}-P_0 B_{\beta } E_{\Theta _e} L_{\Theta _i}-B_0 P_{\beta } E_{\Theta _i} L_{\Theta _e}+B_0 E_{\Theta _e} P_{\beta } L_{\Theta _i}-L_0 B_{\beta } E_{\Theta _i} P_{\Theta _e}+B_0 L_{\beta } E_{\Theta _i} P_{\Theta _e}+E_0 B_{\beta } P_{\Theta _e} L_{\Theta _i}-B_0 E_{\beta } P_{\Theta _e} L_{\Theta _i}\\
&+B_{\Theta _i} \left(L_{\Theta _e} \left(E_0 P_{\beta }-P_0 E_{\beta }\right)+E_{\Theta _e} \left(P_0 L_{\beta }-L_0 P_{\beta }\right)+\left(L_0 E_{\beta }-E_0 L_{\beta }\right) P_{\Theta _e}\right)+P_{\Theta _i} \left(E_{\Theta _e} \left(L_0 B_{\beta }-B_0 L_{\beta }\right)+\left(B_0 E_{\beta }-E_0 B_{\beta }\right) L_{\Theta _e}\right)\\
&+B_{\Theta _e} \left(E_{\Theta _i} \left(L_0 P_{\beta }-P_0 L_{\beta }\right)+\left(P_0 E_{\beta }-E_0 P_{\beta }\right) L_{\Theta _i}+\left(E_0 L_{\beta }-L_0 E_{\beta }\right) P_{\Theta _i}\right),\\
{\cal N}_{4u}=&-B_{\lambda } P_u E_{\Theta _i} L_{\Theta _e}+B_{\lambda } E_{\Theta _e} P_u L_{\Theta _i}+B_u P_{\lambda } E_{\Theta _i} L_{\Theta _e}-B_u E_{\Theta _e} P_{\lambda } L_{\Theta _i}+B_{\lambda } L_u E_{\Theta _i} P_{\Theta _e}-B_u L_{\lambda } E_{\Theta _i} P_{\Theta _e}-B_{\lambda } E_u P_{\Theta _e} L_{\Theta _i}+B_u E_{\lambda } P_{\Theta _e} L_{\Theta _i}\\
&+B_{\Theta _i} \left(E_{\Theta _e} \left(L_u P_{\lambda }-L_{\lambda } P_u\right)+L_{\Theta _e} \left(E_{\lambda } P_u-E_u P_{\lambda }\right)+P_{\Theta _e} \left(E_u L_{\lambda }-E_{\lambda } L_u\right)\right)+P_{\Theta _i} \left(E_{\Theta _e} \left(B_u L_{\lambda }-B_{\lambda } L_u\right)+L_{\Theta _e} \left(B_{\lambda } E_u-B_u E_{\lambda }\right)\right)\\
&+B_{\Theta _e} \left(L_{\Theta _i} \left(E_u P_{\lambda }-E_{\lambda } P_u\right)+E_{\Theta _i} \left(L_{\lambda } P_u-L_u P_{\lambda }\right)+P_{\Theta _i} \left(E_{\lambda } L_u-E_u L_{\lambda }\right)\right),\\
{\cal N}_{40}=&P_0 \left(-B_{\lambda }\right) E_{\Theta _i} L_{\Theta _e}+P_0 B_{\lambda } E_{\Theta _e} L_{\Theta _i}+B_0 P_{\lambda } E_{\Theta _i} L_{\Theta _e}-B_0 E_{\Theta _e} P_{\lambda } L_{\Theta _i}+L_0 B_{\lambda } E_{\Theta _i} P_{\Theta _e}-B_0 L_{\lambda } E_{\Theta _i} P_{\Theta _e}-E_0 B_{\lambda } P_{\Theta _e} L_{\Theta _i}+B_0 E_{\lambda } P_{\Theta _e} L_{\Theta _i}\\
&+B_{\Theta _i} \left(E_{\Theta _e} \left(L_0 P_{\lambda }-P_0 L_{\lambda }\right)+L_{\Theta _e} \left(P_0 E_{\lambda }-E_0 P_{\lambda }\right)+\left(E_0 L_{\lambda }-L_0 E_{\lambda }\right) P_{\Theta _e}\right)+P_{\Theta _i} \left(E_{\Theta _e} \left(B_0 L_{\lambda }-L_0 B_{\lambda }\right)+\left(E_0 B_{\lambda }-B_0 E_{\lambda }\right) L_{\Theta _e}\right)\\
&+B_{\Theta _e} \left(\left(E_0 P_{\lambda }-P_0 E_{\lambda }\right) L_{\Theta _i}+E_{\Theta _i} \left(P_0 L_{\lambda }-L_0 P_{\lambda }\right)+\left(L_0 E_{\lambda }-E_0 L_{\lambda }\right) P_{\Theta _i}\right),\\
{\cal D}_2=&B_{\beta } \left(B_{\beta } L_{\lambda }-B_{\lambda } L_{\beta }\right),\\ 
\end{aligned}$$
Similarly, the coefficients in the equations (23-26) are obtained as,
$$
\Theta_{e11}=\frac{{\cal N}_{1u}}{{\cal D}_1},~~
\Theta_{e12}=\frac{{\cal N}_{10}}{{\cal D}_1},~~
\Theta_{i11}=\frac{{\cal N}_{2u}}{{\cal D}_1},~~
\Theta_{i12}=\frac{{\cal N}_{20}}{{\cal D}_1},~~
\lambda_{11}=\frac{{\cal N}_{3u}}{{\cal D}_1},~~
\lambda_{12}=\frac{{\cal N}_{30}}{{\cal D}_1},~~
\beta_{11}=\frac{{\cal N}_{4u}}{{\cal D}_1},~~{\rm and}~~
\beta_{12}=\frac{{\cal N}_{40}}{{\cal D}_1}.
$$

\label{lastpage}

\begin{thebibliography}{99}
\bibitem[\protect\citeauthoryear{Agarwal et al.}{2012}]{Agarwal-etal2012}
Agarwal S., Das T.~K., Dey R., Nag S., 2012, GReGr, 44, 1637

\bibitem[\protect\citeauthoryear{Aktar et al.}{2017}]{Aktar-etal2017}
Aktar R., Das S., Nandi A., Sreehari H., 2017, \mnras, 471, 4806

\bibitem[{{Artemova}, {Bjoernsson} \& {Novikov}(1996){Artemova}, {Bjoernsson},  \& {Novikov}}]{Artemova-etal1996}
{Artemova} I.~V., {Bjoernsson} G., {Novikov} I.~D., 1996, \apj, 461, 565

\bibitem[{{Aschenbach}(2010)}]{Aschenbach2010}
{Aschenbach} B., 2010, \memsai, 81, 319

\bibitem[\protect\citeauthoryear{Becker \& Kazanas}{2001}]{Becker-Kazanas01}
Becker P. A., Kazanas D., 2001, \apj, 546, 429

\bibitem[\protect\citeauthoryear{Belloni et al.}{2006}]
{Belloni-etal2006} Belloni T., et al., 2006, MNRAS, 367, 1113 

\bibitem[\protect\citeauthoryear{Belloni}{2010}]
{Belloni2010} Belloni T.~M., 2010, AIPC, 1248, 107 

\bibitem[\protect\citeauthoryear{Belloni, Motta, \& Mu{\~n}oz-Darias}{2011}]
{Belloni-etal2011} Belloni T.~M., Motta S.~E., Mu{\~n}oz-Darias T., 2011, BASI, 39, 409 

\bibitem[\protect\citeauthoryear{Chakrabarti}{1989}]{Chakrabarti1989}
Chakrabarti S.~K., 1989, ApJ, 347, 365 

\bibitem[\protect\citeauthoryear{Chakrabarti \& Wiita}{1992}]{Chakrabarti-Wiita1992} 
Chakrabarti S.~K., Wiita P.~J., 1992, ApJ, 387, L21

\bibitem[\protect\citeauthoryear{Chakrabarti \& Khanna}{1992}]
{Chakrabarti-Khanna1992} Chakrabarti S.~K., Khanna R., 1992, MNRAS, 256, 300 

\bibitem[{Chakrabarti \& Molteni(1995)}]{Chakrabarti-Molteni1995}
Chakrabarti S.~K., Molteni D., 1995, Monthly Notices of the Royal Astronomical
  Society, 272, 80

\bibitem[\protect\citeauthoryear{Chakrabarti \& Titarchuk}{1995}]
{Chakrabarti-Titarchuk1995} Chakrabarti S., Titarchuk L.~G., 1995, ApJ, 455, 623 

\bibitem[\protect\citeauthoryear{Chakrabarti \& Das}{2004}]
{Chakrabarti-Das2004} Chakrabarti S.~K., Das S., 2004, MNRAS, 349, 649

\bibitem[\protect\citeauthoryear{Chakrabarti \& Mandal}{2006}]
{Chakrabarti-Mandal2006} Chakrabarti S.~K., Mandal S., 2006, ApJL, 642, L49

\bibitem[{Chakrabarti \& Mondal(2006)}]{Chakrabarti-Mondal2006}
Chakrabarti S.~K., Mondal S., 2006, Monthly Notices of the Royal Astronomical
  Society, 369, 976

\bibitem[{{Chandrasekhar}(1939)}]{Chandrasekhar1939}
{Chandrasekhar} S., 1939, {An introduction to the study of stellar structure}

\bibitem[\protect\citeauthoryear{Chael, Narayan \& Johnson}{2019}]{Chael-etal2019}
Chael A., Narayan R., Johnson M.~D., 2019, MNRAS, 486, 2873


\bibitem[{Colpi, Maraschi \& Treves(1984)Colpi, Maraschi, \&  Treves}]{Colpi-etal1984}
Colpi M., Maraschi L., Treves A., 1984, The Astrophysical Journal, 280, 319

\bibitem[\protect\citeauthoryear{Das, Chattopadhyay \& Chakrabarti}{2001}]{Das-etal2001a}
Das S., Chattopadhyay I., Chakrabarti S.~K., 2001, ApJ, 557, 983

\bibitem[\protect\citeauthoryear{Das, et al.}{2001}]{Das-etal2001}
Das S., Chattopadhyay I., Nandi A., Chakrabarti S.~K., 2001, A\&A, 379, 683

\bibitem[{Das(2007)}]{Das2007}
Das S., 2007, Monthly Notices of the Royal Astronomical Society, 376, 1659

\bibitem[\protect\citeauthoryear{Das, et al.}{2014}]{Das-etal2014} 
Das S., Chattopadhyay I., Nandi A., Molteni D., 2014, MNRAS, 442, 251


\bibitem[\protect\citeauthoryear{Das, Pendharkar \& Mitra}{2003}]{Das-etal2003}
Das T.~K., Pendharkar J.~K., Mitra S., 2003, ApJ, 592, 1078

\bibitem[\protect\citeauthoryear{Das \& Czerny}{2012}]
{Das-Czerny2012} Das T.~K., Czerny B., 2012, NewA, 17, 254 

\bibitem[\protect\citeauthoryear{Dihingia, Das, \& Mandal}{2015}]
{Dihingia-etal2015} Dihingia I.~K., Das S., Mandal S., 2015, ASInC, 12,  

\bibitem[\protect\citeauthoryear{Dihingia, Das, \& Mandal}{2018a}]{Dihingia-etal2018a} 
Dihingia I.~K., Das S., Mandal S., 2018, MNRAS, 475, 2164 
  
\bibitem[\protect\citeauthoryear{Dihingia, Das, \& Mandal}{2018b}]
{Dihingia-etal2018b} Dihingia I.~K., Das S., Mandal S., 2018, JApA, 39, 6 

\bibitem[\protect\citeauthoryear{Dihingia et al.}{2018c}]{Dihingia-etal2018}
Dihingia I.~K., Das S., Maity D., Chakrabarti S., 2018, PhRvD, 98, 083004 

\bibitem[\protect\citeauthoryear{Dihingia, Das \& Nandi}{2019}]
{Dihingia-etal2019a} Dihingia I.~K., Das S., Nandi A., 2019, MNRAS, 484, 3209

\bibitem[\protect\citeauthoryear{Dihingia, et al.}{2019}]
{Dihingia-etal2019b} Dihingia I.~K., Das S., Maity D., Nandi A., 2019, MNRAS, 488, 2412

\bibitem[\protect\citeauthoryear{Dunn, Fender, K{\"o}rding, Cabanac \& Belloni}{2008}]{Dunn-etal2008}
 Dunn R.~J.~H., Fender R.~P., K{\"o}rding E.~G., Cabanac C., Belloni T., 2008, MNRAS, 387, 545

\bibitem[\protect\citeauthoryear{Esin, McClintock \& Narayan}{1997}]{Esin-etal1997} 
Esin A.~A., McClintock J.~E., Narayan R., 1997, ApJ, 489, 865

\bibitem[\protect\citeauthoryear{Frank, King, \& Raine}{2002}]
{Frank-etal2002} Frank J., King A., Raine D.~J., 2002, Cambridge University Press 

\bibitem[{{Fukue}(1987)}]{Fukue1987}
{Fukue} J., 1987, \pasj, 39, 309

\bibitem[\protect\citeauthoryear{Fukumura \& Tsuruta}{2004}]
{Fukumura-Tsuruta2004} Fukumura K., Tsuruta S., 2004, ApJ, 611, 964

\bibitem[\protect\citeauthoryear{Fukumura \& Kazanas}{2007}]
{Fukumura-Kazanas2007} Fukumura K., Kazanas D., 2007, ApJ, 669, 85 


\bibitem[\protect\citeauthoryear{Fukumura et al.}{2016}]{Fukumura-etal2016}
Fukumura K., Hendry D., Clark P., Tombesi F., Takahashi M., 2016, ApJ, 827, 31


\bibitem[\protect\citeauthoryear{Gierli{\'n}ski \& Newton}{2006}]
{Gierlinski-Newton2006} Gierli{\'n}ski M., Newton J., 2006, MNRAS, 370, 837 

\bibitem[\protect\citeauthoryear{Gou et al.}{2009}]{Gou-etal2009}
Gou L., et al., 2009, ApJ, 701, 1076 

\bibitem[\protect\citeauthoryear{Gou et al.}{2011}]
{Gou-etal2011} Gou L., et al., 2011, ApJ, 742, 85 

\bibitem[\protect\citeauthoryear{Gu \& Lu}{2002}]
{Gu-Lu2002} Gu W.-m., Lu J.-f., 2002, ChA\&A, 26, 147 

\bibitem[\protect\citeauthoryear{Hirose, Krolik, \& Stone}{2006}]
{Hirose-etal2006} Hirose S., Krolik J.~H., Stone J.~M., 2006, ApJ, 640, 901 

\bibitem[\protect\citeauthoryear{Homan \& Belloni}{2005}]
{Homan-Belloni2005} Homan J., Belloni T., 2005, Ap\&SS, 300, 107 

\bibitem[\protect\citeauthoryear{Ivanov \& Prodanov}{2005}]
{Ivanov-Prodanov2005} Ivanov R.~I., Prodanov E.~M., 2005, PhLB, 611, 34 

\bibitem[\protect\citeauthoryear{Kato, et al.}{1993}]{Kato-etal1993}
Kato S., Wu X.-B., Yang L.-T., Yang Z.-L., 1993, MNRAS, 260, 317

\bibitem[\protect\citeauthoryear{Kim et al.}{2018}]{Kim-2018}
Kim J., Garain S.~K., Chakrabarti S.~K., Balsara D.~S., 2018, MNRAS, 482, 3636

\bibitem[\protect\citeauthoryear{Klein-Wolt et al.}{2002}]
{Klein-Wolt-etal2002} Klein-Wolt M., Fender R.~P., Pooley G.~G., Belloni T., 
Migliari S., Morgan E.~H., van der Klis M., 2002, MNRAS, 331, 745 

\bibitem[\protect\citeauthoryear{Kubota \& Done}{2004}]
{Kubota-Done2004} Kubota A., Done C., 2004, MNRAS, 353, 980 

\bibitem[\protect\citeauthoryear{Kumar, et al.}{2013}]{Kumar-etal2013} Kumar R., Singh C.~B., Chattopadhyay I., Chakrabarti S.~K., 2013, MNRAS, 436, 2864

\bibitem[\protect\citeauthoryear{Laor \& Netzer}{1989}]
{Laor-Netzer1989}Laor A., Netzer H., 1989, MNRAS, 238, 897 

\bibitem[\protect\citeauthoryear{Le \& Becker}{2005}]
{Le-Becker2005} Le T., Becker P.~A., 2005, ApJ, 632, 476 

\bibitem[\protect\citeauthoryear{Le et al.}{2016}]
{Le-etal2016} Le T., Wood K.~S., Wolff M.~T., Becker P.~A., Putney J., 2016, ApJ, 819, 112 

\bibitem[\protect\citeauthoryear{Liu et al.}{2010}]{Liu_etal2010}
Liu J., McClintock J.~E., Narayan R., Davis S.~W., Orosz J.~A., 2010, ApJ, 719, L109 	

\bibitem[\protect\citeauthoryear{Ludlam, Miller, \& Cackett}{2015}]
{Ludlam-etal15} Ludlam R.~M., Miller J.~M., Cackett E.~M., 2015, ApJ, 806, 262

\bibitem[\protect\citeauthoryear{Lu, Gu, \& Yuan}{1999}]
{Lu-etal1999} Lu J.-F., Gu W.-M., Yuan F., 1999, ApJ, 523, 340

\bibitem[\protect\citeauthoryear{Machida, Nakamura, \& Matsumoto}{2006}]
{Machida-etal2006} Machida M., Nakamura K.~E., Matsumoto R., 2006, PASJ, 58, 193 

\bibitem[\protect\citeauthoryear{Madau}{1988}]
{Madau1988} Madau P., 1988, ApJ, 327, 116 

\bibitem[\protect\citeauthoryear{Malkan \& Sargent}{1982}]
{Malkan-Sarget1982} Malkan M.~A., Sargent W.~L.~W., 1982, ApJ, 254, 22 

\bibitem[\protect\citeauthoryear{Malkan}{1983}]
{Malkan1983} Malkan M.~A., 1983, ApJ, 268, 582 

\bibitem[\protect\citeauthoryear{Mandal \& Chakrabarti}{2005}]
{Mandal-Chakrabarti2005} Mandal S., Chakrabarti S.~K., 2005, A\&A, 434, 839 

\bibitem[\protect\citeauthoryear{Manmoto, Mineshige, \& Kusunose}{1997}]{Manmoto-etal1997} 
Manmoto T., Mineshige S., Kusunose M., 1997, ApJ, 489, 791 

\bibitem[\protect\citeauthoryear{Meyer-Hofmeister}{2004}]
{Meyer-Hofmeister2004} Meyer-Hofmeister E., 2004, A\&A, 423, 321 

\bibitem[\protect\citeauthoryear{Mignone, Plewa, \& Bodo}{2005}]
{Mignone-etal2005} Mignone A., Plewa T., Bodo G., 2005, ApJS, 160, 199 

\bibitem[\protect\citeauthoryear{Miyakawa et al.}{2008}]
{Miyakawa-etal2008} Miyakawa T., Yamaoka K., Homan J., Saito K., Dotani T., Yoshida A., Inoue H., 2008, PASJ, 60, 637 

\bibitem[\protect\citeauthoryear{Molteni, Ryu \& Chakrabarti}{1996}]{Molteni-etal96}
 Molteni D., Ryu D., Chakrabarti S.~K., 1996, ApJ, 470, 460

\bibitem[\protect\citeauthoryear{Mukhopadhyay}{2002}]{Mukhopadhyay2002}
Mukhopadhyay B., 2002, ApJ, 581, 427

\bibitem[\protect\citeauthoryear{Mukhopadhyay}{2008}]
{Mukhopadhyay2008} Mukhopadhyay B., 2008, ASSP, 12, 261

\bibitem[\protect\citeauthoryear{Nandi, et al.}{2012}]{Nandi-etal2012}Nandi A., Debnath D., Mandal S., Chakrabarti S.~K., 2012, A\&A, 542, 56

\bibitem[\protect\citeauthoryear{Narayan \& Yi}{1995}]
{Narayan-Yi1995} Narayan R., Yi I., 1995, ApJ, 452, 710 

\bibitem[\protect\citeauthoryear{Narayan, McClintock \& Yi}{1996}]{Narayan-etal1996}
Narayan R., McClintock J.~E., Yi I., 1996, ApJ, 457, 821

\bibitem[\protect\citeauthoryear{Nakamura, et al.}{1997}]{Nakamura-etal1997} 
 	Nakamura K.~E., Kusunose M., Matsumoto R., Kato S., 1997, PASJ, 49, 503

\bibitem[\protect\citeauthoryear{Nishikawa et al.}{2005}]{Nishikawa-2005}
Nishikawa K.-I., Richardson G., Koide S., Shibata K., Kudoh T., Hardee P., Fishman G.~J., 2005, ApJ, 625, 60

\bibitem[\protect\citeauthoryear{Obst, et al.}{2013}]{Obst-etal2013} 
Obst M., et al., 2013, AAS/High Energy Astrophysics Division \#13, 126.54

\bibitem[\protect\citeauthoryear{Oda et al.}{2007}]
{Oda-etal2007} Oda H., Machida M., Nakamura K.~E., Matsumoto R., 2007, PASJ, 59, 457 

\bibitem[\protect\citeauthoryear{Oda et al.}{2010}]
{Oda-etal2010} Oda H., Machida M., Nakamura K.~E., Matsumoto R., 2010, ApJ, 712, 639 

\bibitem[\protect\citeauthoryear{Oda et al.}{2012}]{Oda-etal2012} 
Oda H., Machida M., Nakamura K.~E., Matsumoto R., Narayan R., 2012, PASJ, 64, 15 

\bibitem[\protect\citeauthoryear{Okuda \& Das}{2015}]{Okuda-Das2015}
Okuda T., Das S., 2015, MNRAS, 453, 147

\bibitem[\protect\citeauthoryear{Okuda, et al.}{2019}]{Okuda-etal2019}
Okuda T., Singh C.~B., Das S., Aktar R., Nandi A., Dal Pino E.~M. de G., 2019, PASJ, 71, 49

\bibitem[\protect\citeauthoryear{Peitz \& Appl}{1997}]{Peitz_Appl1997} 
Peitz J., Appl S., 1997, MNRAS, 286, 681 

\bibitem[\protect\citeauthoryear{Pu et al.}{2012}]
{Pu-etal2012} Pu H.-Y., Maity I., Das T.~K., Chang H.-K., 2012, CQGra, 29, 245020 

\bibitem[\protect\citeauthoryear{Quataert \& Narayan}{1999}]{Quataert-Narayan1999} Quataert E., Narayan R., 1999, ApJ, 520, 298

\bibitem[\protect\citeauthoryear{Rajesh \& Mukhopadhyay}{2010}]
{Rajesh-Mukhopadhyay2010} Rajesh S.~R., Mukhopadhyay B., 2010, MNRAS, 402, 961 

\bibitem[\protect\citeauthoryear{Ryan, et al.}{2018}]{Ryan-etal2018}
Ryan B.~R., Ressler S.~M., Dolence J.~C., Gammie C., Quataert E., 2018, ApJ, 864, 126

\bibitem[\protect\citeauthoryear{Ressler, et al.}{2015}]{Ressler-etal2015}
Ressler S.~M., Tchekhovskoy A., Quataert E., Chandra M., Gammie C.~F., 2015, MNRAS, 454, 1848

\bibitem[\protect\citeauthoryear{Riffert \& Herold}{1995}]
{Riffert-Herold1995} Riffert H., Herold H., 1995, ApJ, 450, 508 

\bibitem[\protect\citeauthoryear{Rossi et al.}{2004}]
{Rossi-etal2004} Rossi S., Homan J., Miller J.~M., Belloni T., 2004, NuPhS, 132, 416 

\bibitem[\protect\citeauthoryear{Rybicki \& Lightman}{1979}]
{Rybicki-Lightman1979} Rybicki G.~B., Lightman A.~P., 1979, Radiative processes in astrophysics, Wiley-Interscience

\bibitem[\protect\citeauthoryear{Ryu, Chakrabarti, \& Molteni}{1997}]{Ryu-etal1997}
Ryu D., Chakrabarti S.~K., Molteni D., 1997, \apj, 474, 378

\bibitem[\protect\citeauthoryear{Ryu, Chattopadhyay, \& Choi}{2006}]
{Ryu-etal2006} Ryu D., Chattopadhyay I., Choi E., 2006, ApJS, 166, 410 

\bibitem[\protect\citeauthoryear{S{\k{a}}dowski, et al.}{2017}]
	{Sadowski-etal2017} S{\k{a}}dowski A., Wielgus M., Narayan R., Abarca D., McKinney J.~C., Chael A., 2017, MNRAS, 466, 705

\bibitem[\protect\citeauthoryear{Sarkar \& Das}{2016}]{Sarkar-Das16}
Sarkar B., Das S., 2016, \mnras, 461, 190


\bibitem[\protect\citeauthoryear{Sarkar \& Chattopadhyay}{2019}]
	{Sarkar-Chattopadhyay2019} Sarkar S., Chattopadhyay I., 2019, IJMPD, 28, 1950037


\bibitem[\protect\citeauthoryear{Semer{\'a}k \& Karas}{1999}]
{Semerak-Karas1999}Semer{\'a}k O., Karas V., 1999, A\&A, 343, 325 	

\bibitem[\protect\citeauthoryear{Shakura \& Sunyaev}{1973}]
{Shakura-Sunyaev1973} Shakura N.~I., Sunyaev R.~A., 1973, A\&A, 24, 337 

\bibitem[\protect\citeauthoryear{Shafee et al.}{2006}]{Shafee-etal06} 
Shafee R., McClintock J.~E., Narayan R., Davis S.~W., Li L.-X., Remillard R.~A., 2006, ApJ, 636, L113

\bibitem[\protect\citeauthoryear{Shields}{1978}]{Shields1978} 
Shields G.~A., 1978, Natur, 272, 706 

\bibitem[\protect\citeauthoryear{Sinha, Rajesh, \& Mukhopadhyay}{2009}]
{Sinha-etal2009} Sinha M., Rajesh S.~R., Mukhopadhyay B., 2009, RAA, 9, 1331 
 
\bibitem[\protect\citeauthoryear{Stepney \& Guilbert}{1983}]
{Stepney-Guilbert1983} Stepney S., Guilbert P.~W., 1983, MNRAS, 204, 1269 

\bibitem[\protect\citeauthoryear{Sukov{\'a} \& Janiuk}{2015}]{Sukova-Janiuk2015} Sukov{\'a} P., Janiuk A., 2015, MNRAS, 447, 1565

\bibitem[\protect\citeauthoryear{Sun \& Malkan}{1989}]{Sun-Malkan1989} 
Sun W.-H., Malkan M.~A., 1989, ApJ, 346, 68 

\bibitem[\protect\citeauthoryear{Synge}{1957}]
{Synge1957} Synge J.~L., 1957, The relativistic gas, Vol.~32. North-Holland Amsterdam

\bibitem[\protect\citeauthoryear{Taub}{1948}]
{Taub1948} Taub A., 1948, Physical Review, 74, 328

\bibitem[\protect\citeauthoryear{Takahashi}{2007}]{Takahashi2007} Takahashi R., 2007, A\&A, 461, 393 

\bibitem[\protect\citeauthoryear{Xiao, Li, Yan, Lu, Chen \& Qu}{2018}]{Xiao-etal2018} 
Xiao G.-. cheng ., Li Z.-. jian ., Yan L.-. li ., Lu Y., Chen L., Qu J.-L., 2018, ChA\&A, 42, 48

\bibitem[\protect\citeauthoryear{Yang \& Kafatos}{1995}]{Yang-Kafatos1995}
Yang R., Kafatos M., 1995, A\&A, 295, 238

\bibitem[\protect\citeauthoryear{Yuan, Quataert \& Narayan}{2003}]
	{Yuan-etal2003} Yuan F., Quataert E., Narayan R., 2003, ApJ, 598, 301

\bibitem[\protect\citeauthoryear{Zdziarski et al.}{2004}]
{Zdziarski-etal2004} Zdziarski A.~A., Gierli{\'n}ski M., Miko{\l}ajewska J.,
 Wardzi{\'n}ski G., Smith D.~M., Harmon B.~A., Kitamoto S., 2004, MNRAS, 351, 791 

\end{thebibliography}
\end{document}